\documentclass[lettersize,journal]{IEEEtran}
\usepackage{makecell}
\usepackage{enumitem}
\usepackage{algorithmic}
\usepackage{graphicx}
\usepackage{amsmath}
\usepackage{cite}
\usepackage{amsmath,amssymb,amsfonts}
\usepackage{textcomp}
\usepackage{xcolor}
\usepackage{multirow}
\usepackage{amsthm}
\usepackage{amsfonts}
\usepackage{booktabs}
\usepackage{algorithm}
\usepackage{array}
\usepackage[caption=false,font=normalsize,labelfont=sf,textfont=sf]{subfig}
\usepackage{textcomp}
\usepackage{stfloats}
\usepackage{url}
\usepackage{verbatim}
\usepackage{soul, color, xcolor}
\usepackage[T1]{fontenc}

\usepackage{subcaption} 

\begin{document}

\title{SMTFL: Secure Model Training to Untrusted Participants in Federated Learning}

\author{Zhihui Zhao, Xiaorong Dong, Yimo Ren, Jianhua Wang, Dan Yu, Hongsong Zhu and Yongle Chen% <-this % stops a space

\IEEEcompsocitemizethanks{\IEEEcompsocthanksitem Z. Zhao, X. Dong, J. Wang, D. Yu,  Y. Chen are with the College of Computer Science and Technology, Taiyuan University of Technology, Taiyuan, China, 030024. (E-mail: \{zhaozhihui, wangjianhua02, chenyongle\}@tyut.edu.cn, dongxiaorong0607@163.com)

D. Yu is with the College of Artificial Intelligence, Taiyuan University of Technology, Taiyuan, China, 030024. (E-mail: yudan@tyut.edu.cn)

Y. Ren, H. Zhu are with the Institute of Information Engineering, Chinese Academy of Sciences, and the School of Cyberspace Security, University of Chinese Academy of Sciences. Beijing, China, 100085. (\{renyimo, zhuhongsong\}@iie.ac.cn)\protect\\
\IEEEcompsocthanksitem 
Yongle Chen is the corresponding author.}% <-this % stops an unwanted space
\thanks{Manuscript received x xx, xxxx; revised x xx, xxxx.}}

% \markboth{IEEE Transactions on Mobile Computing, ~Vol.~x, No.~x, x~x}%
%{Shell \MakeLowercase{\textit{et al.}}: Bare Demo of IEEEtran.cls for Computer Society Journals}

\IEEEtitleabstractindextext{%
\begin{abstract}
Federated learning is an essential distributed model training technique. However, threats such as gradient inversion attacks and poisoning attacks pose significant risks to the privacy of training data and the model correctness. We propose a novel approach called SMTFL to achieve secure model training in federated learning without relying on trusted participants. To safeguard gradients privacy against gradient inversion attacks, clients are dynamically grouped, allowing one client's gradient to be divided to obfuscate the gradients of other clients within the group. This method incorporates checks and balances to reduce the collusion for inferring specific client data. To detect poisoning attacks from malicious clients, we assess the impact of aggregated gradients on the global model's performance, enabling effective identification and exclusion of malicious clients. Each client's gradients are encrypted and stored, with decryption collectively managed by all clients. The detected poisoning gradients are invalidated from the global model through a  unlearning method. % To our best knowledge, we present the first practical secure aggregation scheme, 
We present a practical secure aggregation scheme, which does not require trusted participants, avoids the performance degradation associated with traditional noise-injection, and aviods complex cryptographic operations during gradient aggregation. Evaluation results are encouraging based on four datasets and two models: SMTFL is effective against poisoning attacks and gradient inversion  attacks, achieving an accuracy rate of over 95\% in locating malicious clients, while keeping the false positive rate for honest clients within 5\%. The model accuracy is also nearly restored to its pre-attack state when SMTFL is deployed.

\end{abstract}

}
\maketitle

\IEEEdisplaynontitleabstractindextext

\IEEEpeerreviewmaketitle

% \IEEEraisesectionheading{\section{Introduction}\label{sec:introduction}}

\section{Introduction}

\IEEEPARstart{F}{ederated} learning (FL)~\cite{zhang2021survey} is an important technique for multiple clients to jointly train a model with the help of a aggregation server (abbr. as server). Instead of uploading the training data directly, each client uploads its local gradient updates (obtained by training its own data) to the server, where its data privacy is protected. Data is one of the most valuable resources for the model training. With the emergence and development of Large Language Models~\cite{zhao2023survey}, Artificial Intelligence, Edge Intelligence~\cite{zhao2024feashare}, etc., the growing demand for data volume is met by an expanding scale of FL clients and their diverse origins. For example, users, without the abundant data,  launch model training tasks to third parties~\cite{kang2024semi}, third parties use their own data for training and upload gradient updates to server in exchange for the payoff.

\begin{figure}
    \centering
    \includegraphics[width=0.8\linewidth, height = 4cm]{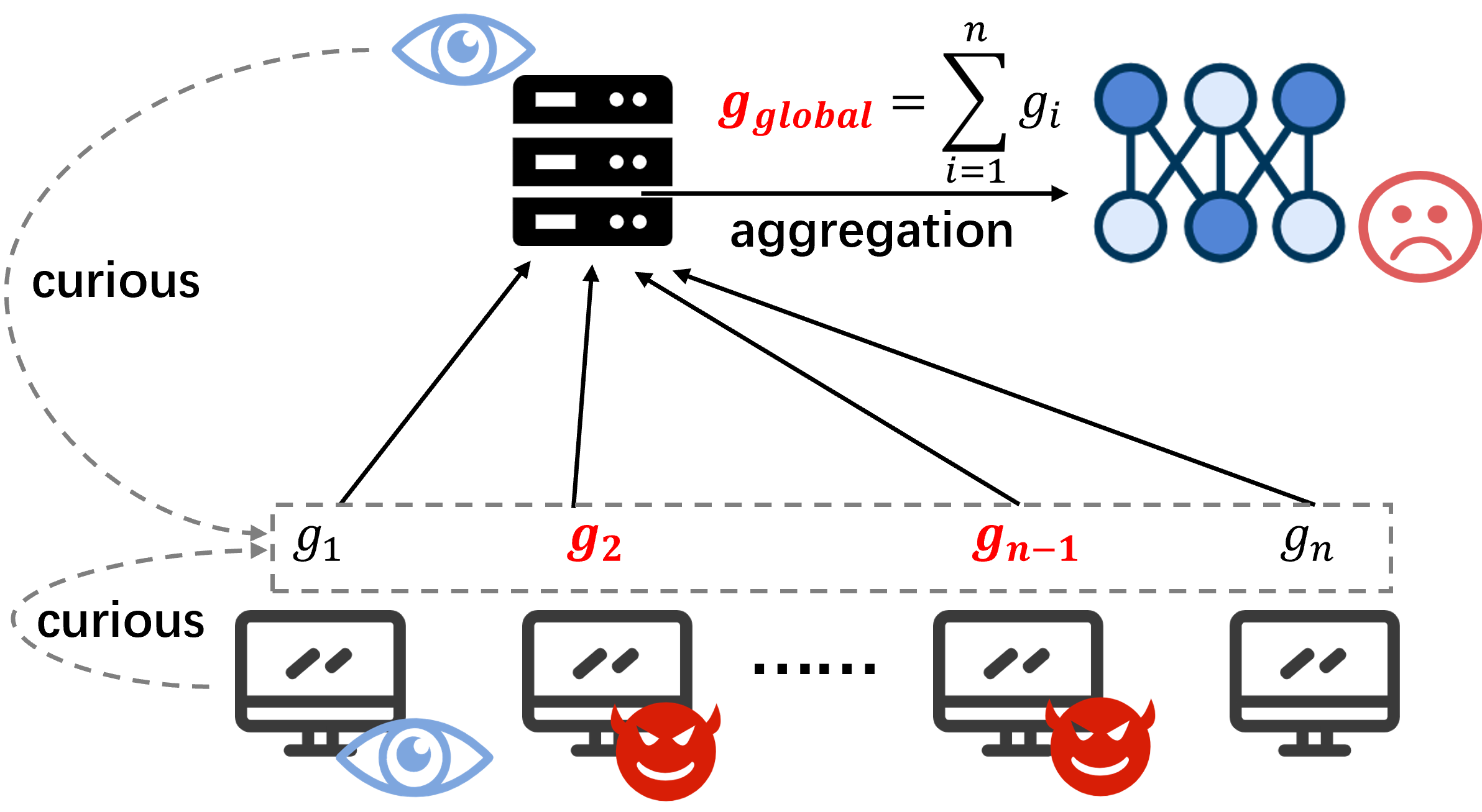}
    \caption{The focused FL security issues in this paper}
    \label{fig1}
\end{figure}

\textbf{Motivation}. Privacy protection of training data and model correctness are critical for the reliable functioning of FL systems. As shown in Fig.\ref{fig1}, FL participants (i.e., clients and server) are untrusted, some may deviate from the established service rules: (1) Model inversion attack~\cite{li2022ressfl}\cite{nguyen2024label}\cite{9996844}. Server and clients may be curious about others' training data, they can potentially recover the specific client's training data based on its gradient updates. (2) Poisoning attack~\cite{wan2024data}\cite{cao2022mpaf}.  Malicious clients upload incorrect gradients, such as inverted gradients, to the server. This action causes the aggregated gradient to deviate from the benign gradient, thereby degrading the performance of FL model.

Some research has been addressing these security issues for FL. On one hand, some work~\cite{9996844, 9474309} assumes that FL participants are semi-honest, i.e., participants perform their tasks correctly, but they attempt to violate the privacy rules and derive clients' training data. Secure aggregation techniques (homomorphic encryption~\cite{10321722}, differential privacy (DP)~\cite{truex2020ldp}, and gradient masking~\cite{kim2024random}, etc.) are widely employed to safeguard data privacy. However, these methods may encounter challenges, including high computational overhead due to complex cryptographic operations, model performance degradation from noisy data insertion, and a focus on the transmission security of gradients rather than its computational security. On the other hand, some studies (FLTrust~\cite{cao2020fltrust}, Ozdayi M S et.al~\cite{ozdayi2021defending}) have attempted to defend against poisoning attacks from malicious clients. They assume that the server is trustworthy and determine whether a client is malicious through collecting and training on a small amount of publicly available data pertaining to that client. According to the majority principle, the gradient direction uploaded by the majority of clients is deemed to be the correct direction. However, the server is semi-honest and can infer the clients' training data via these public data. It may misjudge poisoning attacks due to the data bias and scarcity of public data. Malicious clients may also collude~\cite{lyu2023poisoning, 10387506} to jointly upload gradients with the same but incorrect direction, or there may be multiple collusion groups, making it possible that there is no consensus on the gradient direction among clients. Eiffel~\cite{roy2022eiffel} addresses both input privacy and integrity protection, but it suffers from high complexity, making it challenging to balance model robustness with computational efficiency. Therefore, there are several \textbf{challenges} to address both security issues and the limitations of existing work.

\begin{itemize}
    \item During the interaction and gradient aggregation phase, both clients and the server may obtain a specific client's gradient updates if these gradients are aggregated in an unencrypted state and without added noise for obfuscation. This situation provides malicious participants with the opportunity to infer the training data of clients.
    
    \item In the absence of a trusted participant and without the need for the server to collect public data for each client, it becomes challenging to directly ascertain the correctness of a client's gradient. A single gradient update from a client with a limited data volume is likely to have a minimal impact on the global model's performance. The availability of the majority principle may be compromised by collusion attacks.
    
    \item  If incorrect gradients from malicious clients are not detected promptly, they will be integrated into the global model during the undetected window period. Therefore, this scenario will degrade the long-term performance of the global model.
\end{itemize}

\textbf{Methods}. We propose a novel approach called SMTFL to cope the above challenges:
\begin{itemize}
\item To protect the gradients privacy, we group every three clients, where one client's gradient is dynamically divided into two shares to obfuscate the gradients of the other clients. The gradients from this group are ultimately aggregated by one client and uploaded to the server for getting the global model. To prevent some clients from obtaining a specific client's gradient through collusion attacks, our gradient aggregation rules establish a system of checks and balances among the group clients, such that colluders, in the act of exposing others' gradients, will also reveal their own.

\item To detect poisoning attacks from malicious clients, we evaluate the impact of the aggregated gradients from a group of clients on the global model's performance. We assess this group of clients collectively based on the observed changes in the global model's performance. Although this rule may incidentally affect some honest clients during a single update, the dynamic grouping rule mitigates the overall impact on honest clients across each update session. When a client receives negative evaluations beyond a preset threshold, it will be considered as malicious and excluded from the FL system.

\item To mitigate the impact of the aggregated poisoning gradients on the global model, each client's gradients are encrypted and regularly stored on a storage server. The decryption keys for these gradients are not held by any single entity but are collectively managed by all clients. Upon the detection of a malicious client, its gradients are decrypted under the consensus of the majority of clients. The server then uses the idea of federated unlearning~\cite{liu2024survey} to mitigate the influence of the poisoning gradients.
\end{itemize}

\textbf{Contribution}.  Our contributions are outlined as follows.
\begin{itemize}
    \item %To our best knowledge, 
    SMTFL is a practical and secure model training scheme in FL scenarios without trusted participants. It can preserves gradient privacy against the gradient inversion attacks by servers and clients, and ensures the performance of the global model, safeguarding it from poisoning attacks orchestrated by clients.
    
    \item In the privacy protection, SMTFL does not use the noise, such as that introduced by differential privacy, to avoid any adverse impact on the global model's performance. It eliminates the need for complex cryptographic operations in aggregation. To defense against poisoning attacks, the FL system does not require trusted participants, and the server is not obligated to collect public data for each client beforehand. SMTFL can locate and remove malicious clients. It also mitigates the influence of aggregated poisoning gradients on the global model.

    \item We evaluate SMTFL across four datasets and two models, and the results are encouraging: SMTFL is effective against poisoning attacks and gradient inversion  attacks, achieving an accuracy rate of over 95\% in locating malicious clients, while keeping the false positive rate for honest clients within 5\%. Meanwhile, the model accuracy can be nearly restored to its pre-attack state. Even under varying data distributions and proportions of malicious clients, SMTFL still shows excellent performance. We also demonstrate the advantages of SMTFL by comparing it with related work.
\end{itemize}

The rest is organized as follows. Section~\ref{Pre} and Section~\ref{Threat} describe the preliminaries, threat model, and our security goals, respectively. The SMTFL details are introduced in Section~\ref{App}. Section~\ref{security} and Section~\ref{Exp} present the security analysis and experiment evaluation, respectively. Section~\ref{re} introduces the related work. Section~\ref{Dis} discusses and concludes this paper.

\section{Preliminaries}
\label{Pre}
\subsection{Primary reasons that clients are untrusted}
Clients are untrusted in the FL system. On one hand, clients are vulnerable to external attacks. Any computing-capable device may join the FL system, but individual users may lack the expertise to configure proper security strategies for their smart devices~\cite{zhao2024feashare, zhao2022detection}. Attackers can leverage these compromised clients to reveal others' gradients information or upload incorrect gradient updates. On the other hand, client's administrators may subjectively deviate from established rules. Semi-honest and malicious clients seek to obtain other clients' private training data. Malicious clients also upload incorrect gradient updates to sabotage the FL system (e.g., malicious competition). In some FL applications related to crowdsourced computing~\cite{kang2024semi}, some clients, facing data scarcity, may fabricate false gradients or manipulate gradients trained on public datasets prior to uploading them to the server. Additionally, some profit-driven clients may refrain from training, choosing instead to make minor modifications to the global model received from the server to conserve computational resources. These actions pose security threats to the FL system.

\subsection{Federated unlearning}
Federated Unlearning (FUL)~\cite{liu2024survey} focuses on the removal of specific clients' contributions or data samples from the global model, aiming to safeguard privacy and enhance security. The core lies in adjusting model parameters to achieve an effect equivalent to that of training without specific data, while avoiding complete retraining. Based on FUL, the server can, at user request, make the model forget the contributions trained by specific data. Therefore, server can enhance the model's security by making the model forget updates trained from incorrect or low-quality data. FUL methods include contribution deletion~\cite{zhu2023heterogeneous}, local parameter adjustment~\cite{liu2021revfrf}, training update correction~\cite{gao2024verifi}, and training gradient correction~\cite{halimi2022federated}, etc.

\subsection{Threshold encryption} 
Threshold encryption, a cryptographic technique, enhances the security and reliability of key management. This technique involves splitting a key into multiple shares, with the stipulation that the key can only be recovered when at least $t$ shares are combined, thereby enabling data decryption. The method improves the security and fault tolerance of the key by eliminating single points of failure. Threshold encryption is primarily based on the Threshold Secret Sharing Protocol, such as the Shamir scheme~\cite{yu2021blockchain}.

Specifically, a key $S$ is divided into $N$ key shares $\{S_1, S_2,...,S_N\}$, and a threshold value $t$ is set to indicate that at least $t$ key share unions are required to recover the original key $S$. For instance, consider the key $S$ as a split secret, the Shamir scheme generates a random polynomial:
\begin{center}
$f(x)=S+a_1x+a_2x^2+ ...+a_{t-1}x^{t-1} \quad (\bmod\ p)$
\end{center}
where $S$ is the secret, $a_1, a_2, \dots, a_{t-1}$ are the randomly generated coefficients, $p$ is the prime, and $x$ is the index of the participant (e.g. $x_1,x_2,\dots,x_N$). Each participant computes its key share $(x_i, f(x_i))$.

When at least $t$ participants collaborate, given $t$ key shares $(x_1, y_1)$, $(x_2,y_2)$,$\dots$, $(x_t, y_t)$, the secret $S$ can be recovered using Lagrange interpolation:
\begin{center}
    $S = \sum\limits_{i = 1}^t {{y_i}} \prod\limits_{1 \le j \le t,j \ne i} {\frac{{{x_j}}}{{{x_j} - {x_i}}}} \quad(\bmod\ p)$
\end{center}

\section{Threat model and security goals}
\label{Threat}
In the traditional threat model~\cite{zhao2024feashare}, participants are usually classified into 1) Honest: participants perform their assigned tasks correctly according to the established rules and do not violate the system’s security rules. 2) Semi-honest: participants honestly follow the rules without cheating and other misbehavior, but they attempt to violate the data privacy rules. 3) Malicious: participants intentionally violate the security rules to fail the services and obtain privacy data. There are two primary roles in FL systems: the server and the clients. In this paper, we not only consider the possibility that clients may execute poisoning attack to disrupt model training, but also that clients and server may both execute gradient inversion attacks to obtain the training data from other clients. We make the following \textbf{assumptions} for clients and server, respectively.

\begin{itemize}
    \item \textbf{We assume the server to be semi-honest}. Specifically, it correctly aggregates the gradient uploaded by the clients and returns the correct gradient to the clients participating in this epoch. However, the server is curious about the clients' training data and may attempt to perform gradient inversion attacks. We consider that the assumption is reasonable. The server, typically operated by the system's administrators, is responsible for aggregating the global model. Its goal is to collaborate with numerous clients to train the model. If the server is indifferent to the performance of the model being trained (i.e., whether it is subject to poisoning attacks), it could potentially act as a poisoner itself, potentially undermining the entire training process of the FL system.

    \item \textbf{We assume the clients to be semi-honest or malicious}. Specifically, semi-honest clients correctly compute local gradients based on their own data and participate in the aggregation of global models. However, they may attempt to obtain the training data from other clients through illegal methods, such as gradient inversion attacks. Malicious clients not only seek to acquire the training data from others, but they also try to compromise the global model through poisoning attacks. Malicious clients upload incorrect gradients to the server in each epoch, meaning their rate of malicious activity is 100\%. Furthermore, clients may collude to obtain the gradients of other clients or to compromise the global model (i.e., collusion attack). We notice a fact: all clients are rational and will not collude with other participants for no reason, especially when doing so would harm their interests or no extra benefits.
\end{itemize}

Most studies~\cite{9996844, 9474309} typically assume that clients and server are semi-honest, or that only clients are malicious. To our best knowledge, few works have concurrently addressed both scenarios. This paper considers a much stronger threat model and should meet the following security goals: (1) Both clients and server are unable to get the gradients of specific clients; (2) Detecting and locating malicious clients that compromise FL system; (3) The performance of global model does not significantly deteriorate in the presence of malicious clients.

\begin{figure*}
    \centering
    \includegraphics[width=0.95\linewidth]{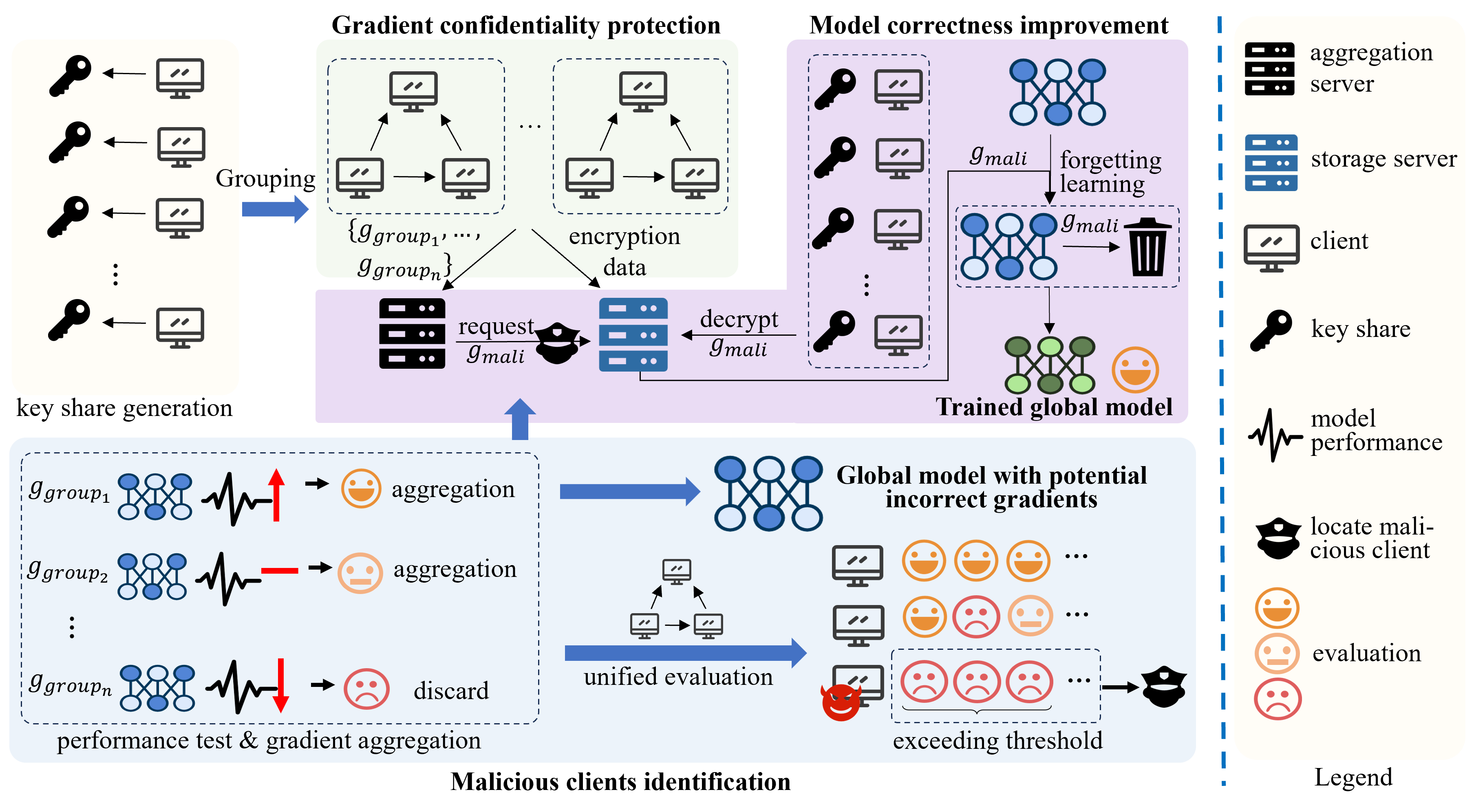}
    \caption{The framework of SMTFL}
    \label{fig2}
\end{figure*}

\textbf{Formalization}. With loss of generality, we consider a FL system comprising clients $\{c_1, ..., c_m\}$ and an aggregation server $s_{a}$. During the $k$-th epoch, $\forall$ $c_i$ utilizes its data $d_{c_i}$ to train and obtain the updated local gradient $g_{i}^{k}$, which is uploaded to the $s_a$. The $s_a$ aggregate all $g_{i}^{k}$ ($i\in[1,m]$) to derive the global gradient $g^k$, which is then distributed to all clients for the $(k+1)$-th epoch. Every $c_i$ and $s_a$ attempt to access the $g_{j}^{k}$ of $c_j$ to infer the data $d_{c_j}$ ($j\in[1,m]$). Furthermore, $\exists$ malicious $c_i$ compromises the global model through uploading incorrect local gradients (e.g., inverting gradient direction). That is, the performance gap exceeds the accuracy threshold $\delta$ between the actual and ideal global model. In this paper, a designed scheme $\Pi$ should meet the following security goals:

\begin{itemize}
    \item \textbf{Gradients confidentiality}. Without authorization, no client or server can obtain the gradient $g_{i}^{k}$ of a specific client $c_i$ ($i\in[1,m]$) during the $k$-th epoch, even if they collaborate in the collusion attack.
    \item \textbf{Identifying malicious clients}. Malicious clients can be identified without the trusted participants. The server does not collect a small amount of public data from each client for training purposes to verify their gradients.
    \item \textbf{Correctness of the global model}. After identifying malicious clients, FL system should mitigate the impact of incorrect gradients on the global model, avoiding to degrade the performance of global model obviously.
\end{itemize}

\begin{table}
\centering
\caption{The main notations in this paper.}\label{tab1}
\begin{tabular}{|p{35pt}|p{190pt}|}
% \begin{tabular}{|l|l|}
\hline
{\textbf{Notations}} & \makecell[c]{\textbf{Meanings}}\\
\hline
{$c_i$, $i\in\{1,…,m\}$} & {The $i$-th client, and there are $m$ clients in the FL system} \\
\hline
\makecell[c]{$s_a$} & {The aggregation server} \\
\hline
\makecell[c]{$ d_{c_i}$} & {The training data of the $c_i$}\\
\hline
\makecell[c]{$group_n$} & {The number of clients in a group.}\\
\hline
\makecell[c]{$g_i$} & {The gradient of the $c_i$.}\\
\hline
\makecell[c]{$\{g_A^1, g_A^2\}$} & {The gradient shares for the $c_A$.}\\
\hline
\makecell[c]{$g_{A,B,C}$} & {The group gradient of the $\{c_A, c_B, c_C\}$.}\\
\hline
\makecell[c]{$\varepsilon$} & {The perturbation negotiated by the client and the server.}\\
\hline
\makecell[c]{$Model_{G}$} & {The global model.}\\
\hline
\makecell[c]{$\tau$} & {The predefined threshold for the allowable performance change of $Model_{G}$.}\\
\hline
\makecell[c]{$eva_i$} & {The evaluation score for the $c_i$.}\\
\hline
\makecell[c]{$eva_i^k$} & {The evaluation score for the $c_i$ in the $k$-th epoch.}\\
\hline
\makecell[c]{$thre_{eva}$} & {The evaluation threshold for determining malicious clients.}\\
\hline
\makecell[c]{$\{P_k, S_k\}$} & {The public and secret keys.}\\
\hline
\makecell[c]{$t$} & {The threshold for Threshold Encryption.}\\
\hline
\makecell[c]{$S_{i}^j$} & {The secret share that client $c_i$ sends client $c_j$.}\\
\hline
\makecell[c]{$d_g$} & {The encrypted data.}\\
\hline
\end{tabular}
\end{table}

\section{Approach design}
\label{App}
We propose an approach called SMTFL to achieve our security goals, as shown in Fig.~\ref{fig2}. To ensure the confidentiality of gradients,We group clients together, using their gradients and gradient shares to conceal each other's gradients. This method allows clients' gradients to protect those of others, mitigating collusion attacks among clients. To identify malicious clients, we assess the impact of the aggregated gradients of a group on the global model, thereby providing a unified evaluation of client behavior. A client that frequently degrades the global model's performance will be classified as malicious. To ensure the correctness of the global model, each client's gradient is encrypted and uploaded to the storage server, where access to and decryption of the gradients require consensus from all clients by providing their respective secret shares. FUL is employed to invalidate the poisoned gradients from the global model, thereby restoring its performance. In this paper, the main notations are summarized in Table.~\ref{tab1}.

\subsection{Gradient confidentiality protection}
\begin{figure}
    \centering
    \includegraphics[width=0.9\linewidth]{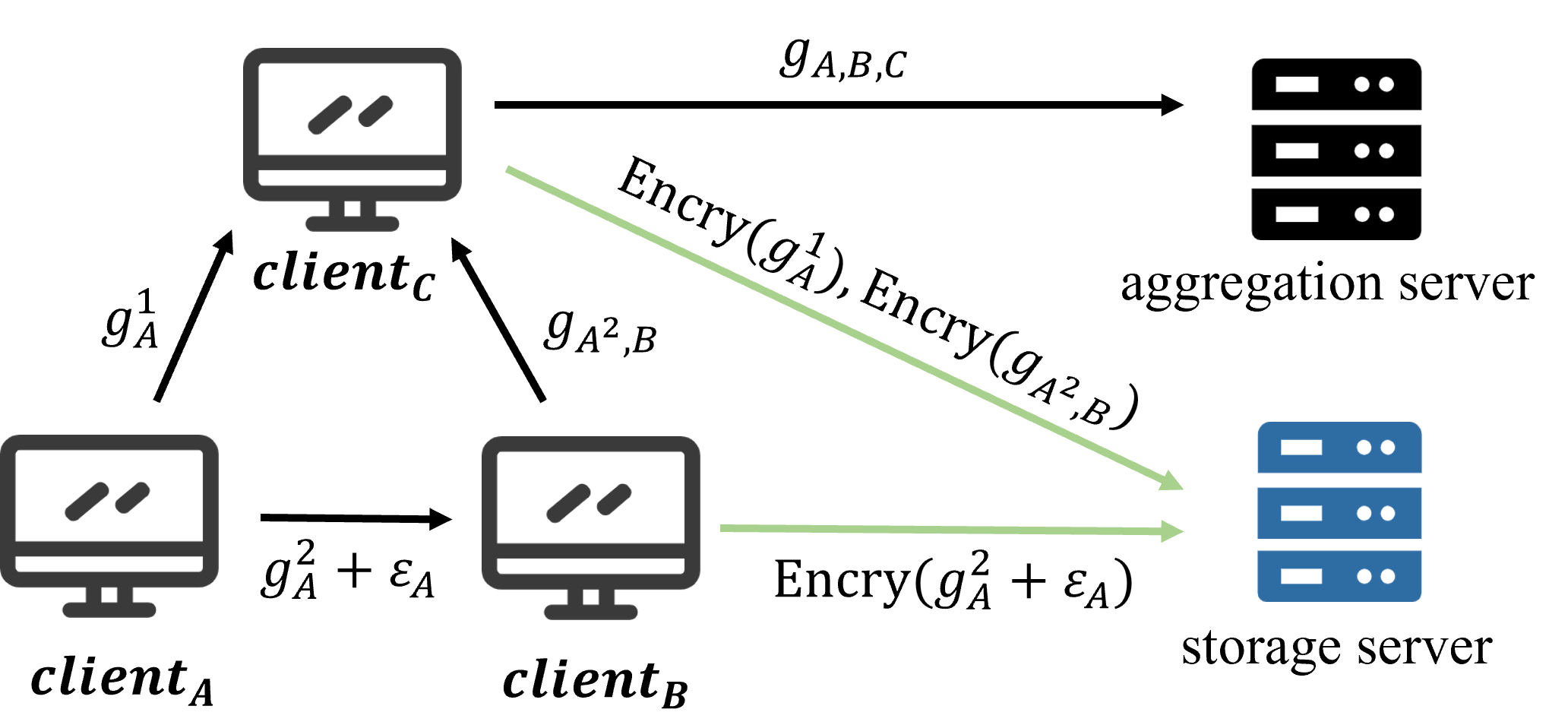}
    \caption{The illustration of gradient aggregation in one group}
    \label{fig3}
\end{figure}

In FL scenarios with untrusted clients and server, where the server is unnecessary to collect a small amount of public data for each client, we aim to reduce the impact of clients' collusion on the global model and identify malicious clients. To achieve this, we group all clients into groups of size $group_n=3$. Based on the idea of edge computing~\cite{shi2016edge}, these gradients from 3 clients are locally aggregated into a group gradient, which is then uploaded to the server by one client. However, untrusted clients and server may leverage gradients to expose the training data of other clients. Therefore, the gradient confidentiality protection must be considered during the group aggregation, as illustrated in Fig. \ref{fig3}.

In protecting gradients, we do not make each client randomly generate noise and add to their local gradient updates (i.e., differential privacy), as this could negatively impact the global model's performance. Nor do we apply homomorphic encryption to the gradients, thereby avoiding complex cryptographic operations. Within a group of clients $\{c_A, c_B, c_C\}$, we randomly split one client's (e.g., $c_A$) gradient:
\begin{equation}
g_A=g_A^1+g_A^2
\end{equation}
where $g_A^1$ is randomly generated, and $\{g_A^1, g_A^2\}$ are distinct from $g_A$. Each client negotiates a dynamic vector $\varepsilon$ with the server upon joining the system (e.g., $c_A$ has the $\varepsilon_A$), which is used to obfuscate their gradients.

Specifically, $c_A$ sends its gradient shares $g_A^1$ and $g_A^2+\varepsilon_A$ to $c_B$ and $c_C$, respectively. $c_B$ obfuscates its gradient $g_B$ by $g_A^2+\varepsilon_A$ and $\varepsilon_B$, and it gets and sends
\begin{equation}
 g_{A^2,B}=g_A^2+\varepsilon_A+g_B+\varepsilon_B
\end{equation}
to $c_C$. The group gradient is aggregated by the $c_C$ as
\begin{equation}
g_{A,B,C}=g_A^1+g_{A^2,B}+ g_C+\varepsilon_C
\end{equation}
Then, $g_{A,B,C}$ is uploaded to the server $s_a$. Since $s_a$ knows $\{\varepsilon_A, \varepsilon_B, \varepsilon_C\}$, it can recover the aggregated precise gradient of this group.

\textbf{After deploying the group aggregation rules, neither any client nor the server can obtain the precise gradients of other clients under normal circumstances.} However, some clients may collude to uncover the precise gradients of other clients. For instance, in the $\{c_A, c_B, c_C, s_a\}$, exposing $c_A$'s gradient $g_A$ requires the collusion among $\{c_B, c_C, s_a\}$. We have observed a fact: during a gradient disclosure operation performed by one party, it may share false gradients to deceive other conspirators. Consequently, three colluding parties will independently disclose $g_A$. In this collusion:
\begin{itemize}
    \item $c_C$ obtains $g_A^2+\varepsilon_A$ and $\varepsilon_A$ from $c_B$ and $s_a$, respectively. 
    \item $c_B$ obtains $g_A^1$ and $\varepsilon_A$ from $c_C$ and $s_a$.
    \item $s_a$ obtains $g_A^1$ and $g_A^2+\varepsilon_A$ from $c_C$ and $c_B$, respectively. 
\end{itemize}
All colluding parties can independently disclose $g_A$. However, in the process of disclosing $g_A$, $c_C$ also possesses 
\begin{center}
$\{g_A^2+\varepsilon_A+g_B+\varepsilon_B,\ g_A^2,\ \varepsilon_A\}$  
\end{center}
$c_C$ could potentially collude again with $s_a$ to obtain $\varepsilon_B$, thereby acquiring $g_B$. Therefore, $c_B$ is unwilling to collude with $c_C$ and $s_a$. The collusion revealing $g_B$ and $g_C$ follows the similar pattern, we will not repeat the description.

% 因此,$client_B$不将是乐意和$client_C$和$s_c$共谋的.同理,在梯度$grad_B$的共谋揭露中, $client_A$需要从$client_C$和$s_c$分别得到$grad_A^2+grad_B+\varepsilon_B$和$\varepsilon_B$. $client_C$需要从$client_A$和$s_c$分别得到$grad_A^2$和$\varepsilon_B$. 此时,$client_C$拥有$grad_A^2$和$grad_A^1+\varepsilon_A$, 则它可以再次与$s_c$共谋得到$\varepsilon_A$, 以揭露$grad_A$.对于$grad_C$的共谋揭露是类似的,故不再进行重复叙述.

In other words, \textbf{three clients impose constraints on one another in one group}. Honest clients will refrain from colluding to reveal other clients' gradients, avoiding their own gradients be exposed as well. It is noteworthy that malicious clients may willingly participate in collusion to reveal the gradients of other clients. Once a client engages in collusion, it indicates that this client has executed a poisoning attack, and its gradients are erroneous or fabricated (i.e., unrelated to its training data). We find that the semi-honest server is a necessary participant in collusion, the server can penalize collusion clients to mitigate the effects of poisoning attacks from malicious clients.

Above, we achieve gradient aggregation and privacy protection among a group of clients locally, alleviating the computational burden on the server. We consider that a group size $group_n=3$ is a reasonable value. On one hand, we find that when the $group_n=2$, a client can easily collude with the server to expose another client's gradient. On the other hand, we want to locate malicious clients in a smaller $group_n$ through monitoring the behaviors of each client. All clients are grouped dynamically in each epoch. If the number of clients falls below 3 in a group, this group will be temporarily excluded from updates to the global model.

\subsection{Malicious clients identification}
In assessing the correctness of gradients, we do not require the server $s_a$ to collect a small amount of public data for each client for training, nor do we rely on the assumption that the gradient direction based on the majority of clients is correct, as this assumption carries the risk of collusion attacks. Instead, for a group of clients $\{c_A, c_B, c_C\}$, we evaluate the impact of their aggregated group gradient $g_{A,B,C}$ on the global model's performance, then \textbf{provide a unified evaluation for their behaviors}. Based on the group gradient, the evaluation method realizes that there is more training data to compute the local gradient updates and to affect the global model more effectively, as shown in the relevant part of Fig.~\ref{fig2}.

We describe the unified evaluation rule for a group of clients. Specifically, for clients $\{c_A, c_B, c_C\}$, each client is assigned an evaluation score $\{eva_A, eva_B, eva_C\}$. Their group gradient $g_{A,B,C}$ is aggregated to the global model $Model_{G}$ in the $k$-th epoch.  Compared to the $(k-1)$-th epoch, there are three possible scenarios:
\begin{itemize}
\item If the performance degradation of $Model_{G}$ exceeds a predefined threshold $\tau$, this is,
\begin{center}
        $Pre_{k-1} - Pre_{k} > \tau$
\end{center}
a poisoning attack event is suspected, where $Pre_{k}$ denotes the performance of the $Model_{G}$ at the $k$-th epoch. Each client is assigned a score $eva_{i}^k=-1$ in the $k$-th epoch, where $i\in\{A, B, C\}$. 

\item If the performance change of $Model_{G}$ falls within an acceptable fluctuation range $[-\tau, \tau]$, that is,
\begin{center}
        $|Pre_{k} - Pre_{k-1}| < \tau$
\end{center}
each client is assigned a score of $eva_{i}^k=0$ in the $k$-th epoch, where $i\in\{A, B, C\}$. This setting allows for a tolerance of non-malicious, occasional computational errors, such as low data quality or natural training fluctuations.

\item If the performance improvement of $Model_{G}$ exceeds the predefined threshold $\tau$, that is, 
\begin{center}
        $Pre_{k} - Pre_{k-1} > \tau$
\end{center}
we consider that no poisoning attack event is detected, and each client is assigned a score of $eva_{i}^k=-1$ in the $k$-th epoch, where $i\in\{A, B, C\}$. 
\end{itemize}
The cumulative evaluation for each client is the sum of the scores assigned over multiple epochs:
\begin{equation}
eva_{i} = \sum eva_{i}^k, i\in\{A, B, C\}
\end{equation}
If the following relationship is met:
\begin{equation}
eva_{i} < thre_{eva}
\end{equation}
In other words, \textbf{the client $c_i$ is considered as malicious if its $eva_{i}$ falls below a preset threshold $thre_{eva}$, and $c_i$ will be removed from the FL system}. Meanwhile, its historical gradients will be invalidated from the $Model_{G}$ to ensure the model's performance (see Section~\ref{App}-C). 

It is noted that while these clients are uniformly scored in the same group, not all clients may be malicious, which could lead to false positives. Assigning a score of $eva_{i}^k=1$ helps mitigate the impact on mistakenly targeted clients. Dynamic client grouping can reduce the effect on such clients. To achieve more precise identification of malicious clients, 3 clients from the same group can be intentionally distributed into different groups in subsequent epochs.

% \hl{In the group gradient, the weight of a client's gradient is related to its evaluation score to encourage honest participants in FL. To simplify the calculation of weights, we set the maximum evaluation score for each client to 0. In aggregating the group gradient, the weights of each client's gradient is positively correlated with its evaluation score, encouraging clients to be honest in FL system. To simplify the calculation of weights, we set the maximum evaluation score to 0. The relationship between the weight $\omega$ and the evaluation score $eva_i$ for the client $c_i$ is set as:}
% \begin{equation}
% \omega = 1 - \rho * |eva_{i}| * In(|eva_{i}|) 
% \end{equation}
% where the $\rho$ is the penalty factor. 

\begin{figure*}
    \centering
    \includegraphics[width=0.9\linewidth]{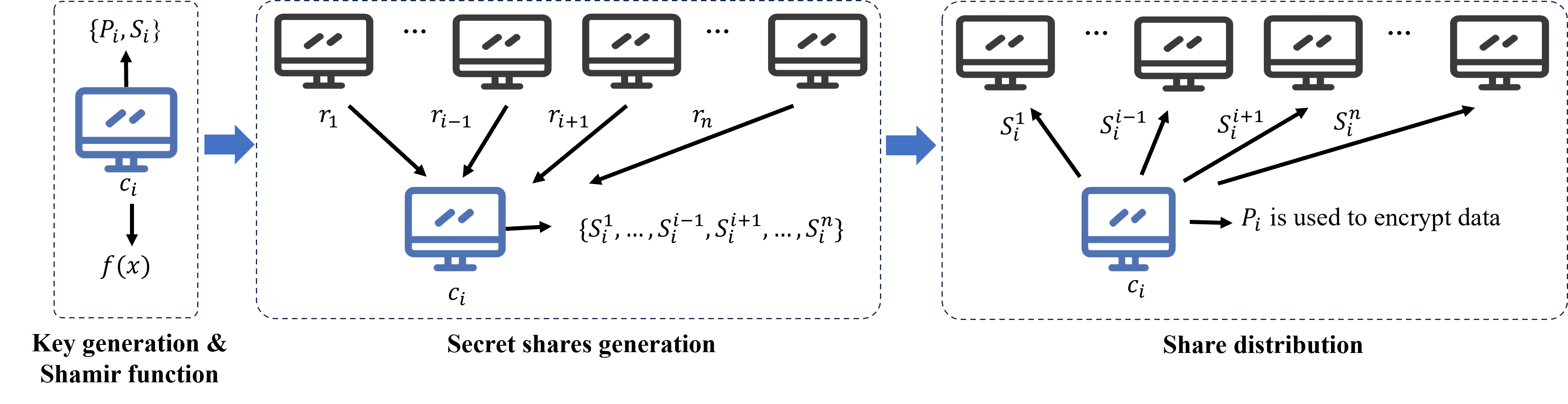}
    \caption{The generation and distribution of secret shares}
    \label{fig5-3}
\end{figure*}

\subsection{Model correctness improvement}
Some incorrect poisoning gradients from malicious clients may be aggregated into the global model for several reasons:
\begin{itemize}
    \item A group consists of both honest and malicious clients. Due to the influence of the honest clients' gradients, their group gradient may not lead to a significant drop in the performance of the global model. 
    \item We evaluate whether a client is malicious based on its cumulative evaluation score. 
\end{itemize}
Therefore, if the performance of global model oscillates within the range $[-\tau, \tau]$ or significantly deteriorates, the $s_a$ will notify the clients in this group to encrypt the received gradients or gradient shares (collectively referred to as gradients) and upload them to a storage server. Once a client is identified to be malicious (see Section~\ref{App}-B), its aggregated historical gradients will be decrypted. Then, we employ the FUL method invalidate these gradients in the global model. The role of the storage server is necessary: 
\begin{itemize}
    \item If each client stores its own gradients, a malicious client might refuse to reveal its poisoning gradients. 
    \item If gradients are stored by clients performing the aggregation, their limited resources may not support storing a large number of historical gradients. 
    \item Clients dynamically join or leave the FL system, potentially preventing the retrieval of poisoning gradients. 
\end{itemize}
The role of storage server could also be played by the $s_a$.

Although gradients are encrypted and stored on the storage server, all participants are untrusted in the FL system, as they all want to access other clients' training data. A critical issue is that the semi-honest server can arbitrarily decrypt and access any client's precise gradient if it independently holds the decryption key for the gradients. This issue seriously compromises the confidentiality of gradients. Therefore, \textbf{we adhere to the idea of multi-party governance, where the decryption key is not held by a single party}. Based on the idea of threshold encryption, the $s_a$ initiates a decryption request, all clients participate in consensus-based decryption. Specifically, it can be divided into three parts: key generation, encryption and decryption of gradients, and gradient forgetting.

In key generation, although the gradients of each client is sent to the storage server, they only need to be confidential from other clients and the storage server. The client will not collude with others to decrypt others' gradients. As shown in Figure.~\ref{fig5-3}, we design that each client $c_i$ $(i\in\{1, ..., m\})$ generates its public-private key pair $\{P_i, S_i\}$. The public key $P_i$ is used to encrypt gradients transmitted to the storage server. The private key $S_i$ forms a threshold encryption scheme $\begin{pmatrix} m \\ t \end{pmatrix}$, and it is divided into $(m-1)$ key shares based on threshold encryption mechanisms:
\begin{center}
    $\{S_i^1, \dots, S_i^{m-1}\}$
\end{center}
At least $t$ shares are required to recover $S_i$. We generate each share $S_i^j$ based on the Shamir scheme, which generates a random polynomial:
\begin{equation}
f(x)=S+a_1x+a_2x^2+ ...+a_{t-1}x^{t-1}\ (\bmod\ p)
\end{equation}

For $\forall$ client $c_j\ (j\in\{1,...,i-1, i+1, ..., m\})$, it sends a random number $r_j$ to the client $c_i$, and $c_i$ inputs the $r_j$ into the $f(x)$ to obtain the $c_j$'s key share $S_i^j$:
\begin{center}
    $S_i^j = (r_j, f(r_j))$
\end{center}
Thus, when the $s_a$ requests to decrypt the detected poisoning gradient $g_{mali}$, at least $t$ clients participate in the decryption, even if a few clients dynamically join or exit the system without providing their shares.

During the process of encrypting gradients and sending them to the storage server, $\forall$ client $c_i$ uses its public key $P_i$ to encrypt the received gradient data $d_g$ (e.g., $g_A^1$, $g_{A^2,B}$ or $g_A^2+\varepsilon_A$):
\begin{equation}
    Data_{decryed}^{c_i} = Encry_{P_i}(d_{g})
\end{equation}
Then, $c_i$ sends $Data_{decryed}^{c_i}$ to the storage server.

After the server $s_a$ has identified the malicious client $c_{mali}$ (see Section~\ref{App}-B), it requests the storage server for the poisoning gradient $g_{mali}$ of $c_{mali}$. 

The storage server requests the key shares from all online clients, and at least $t$ clients send some clients' key shares, where these clients has cooperating with $c_{mali}$. For instance, the $c_i$ has cooperating with $c_{mali}$,  its shares $\{S_i^1,...,S_i^t\}$ is received. The key $S_{i}$ can be recovered using the Lagrange interpolation function:
\begin{equation}
    S_i = \sum\limits_{j = 1}^t {{f(r_j)}} \prod\limits_{1 \le l \le t,l \ne j} {\frac{{{r_l}}}{{{r_l} - {r_j}}}}\ (\bmod\ p)
\end{equation}

Server $s_a$ uses the $S_{i}$ to decrypt the $d_{g}$, that is related $g_{mali}$, from other clients. Meanwhile, the $s_a$ possesses $\varepsilon_{c_{mali}}$ and can obtain the $g_{mali}$.

After the server obtains the poisoning gradients $\{g_{mali}^1, ..., g_{mali}^k\}$ of malicious clients $\{c_{mali}^1, ..., c_{mali}^k\}$, we leverage the FUL to invalidate them in the global model: 
\begin{equation}
    g_{global} = \frac{\sum g_{group} - \sum_{j=1}^{k} g_{mali}^j * \omega_{c_{mali}^j}}{m - k}
\end{equation}
At this point, \textbf{we mitigate the impact of the aggregated poisoning gradients on the global model}.

\section{Security analysis}
\label{security}
In this section, we formally analyze the security of SMTFL. 

\newtheorem{lemma}{Lemma}
\begin{lemma}\label{lemma1}
No client can individually obtain the precise gradients of other clients, thus preventing the acquisition of clients' training data through gradient inversion attacks.
\end{lemma}
\textbf{Proof}. The gradient aggregation is divided into two stages in SMTFL: 1)  Group aggregation among three clients within a group; 2) Aggregation of multiple group gradients on the server. For a group of clients $\{c_A, c_B, c_C\}$,  $c_A$ provides $g_A^2+\varepsilon_A$ and $g_A^1$ to $c_B$ and $c_C$, respectively. $c_B$ provides $g_{A^2,B}$ to $c_C$,  $c_C$ provides $g_{A,B,C}$ to the server $s_a$. Due to the presence of $\varepsilon_A$, $\varepsilon_B$, $\varepsilon_C$, $g_A^1$ and $g_A^2$, no single client or server can independently obtain the precise gradients of $c_A, c_B$ and $c_C$.  In the aggregation of multiple group gradients, each group gradient is formed by the gradients aggregation from three clients. Even if the server possesses $\varepsilon_A$, $\varepsilon_B$ and  $\varepsilon_C$, it still cannot obtain the precise gradients of a specific client.   \hfill $\square$

\providecommand{\lemma}{...} 

\begin{lemma}\label{lemma2}
For a group of clients $\{c_A, c_B, c_C\}$, any two semi-honest clients will not collude with the server $s_a$ to obtain the precise gradient of another client. Meanwhile, the group's clients will not collude to launch a poisoning attack.
\end{lemma}
\textbf{Proof}. As described in Section~\ref{App}-A, we considered the scenario where $c_B$, $c_C$, and $s_a$ collude to obtain $c_A$'s gradient $g_A$. Once $g_A$ is revealed, $c_B$'s gradient $g_B$ will also be revealed by the collusion between $c_C$ and $s_a$. Consequently, semi-honest clients will be unwilling to collude, as their gradients would also be exposed. Regarding poisoning attacks, SMTFL decides whether to aggregate the group gradient based on its impact on the global model. In a group of clients, if two or more clients launch poisoning attacks, the incorrect gradients will dominate the direction of the group's gradients, thus decreasing the global model's performance. Based on the rule in Section~\ref{App}-B, the group's gradients will not be integrated into the global model. Therefore, no client will collude.    \hfill $\square$
\providecommand{\lemma}{...} 
\begin{lemma}\label{lemma3}
Threshold encryption secures the gradients storage from all clients and servers, and it allows for the effective retrieval of poisoning gradients, even when some clients dynamically join or leave the system.
\end{lemma}
\textbf{Proof}. In a system with $m$ clients, each client possesses their own public and private keys. For instance, client $c_{i}$ uses its public key $P_{i}$ to encrypt the received gradient data $Encry_{P_i}(d_g)$.  Its private key $S_i$  is split into $m-1$ shares $S_i^j$ ($j\in\{1,..., i-1, i+1, ..., m\}$) through threshold encryption mechanisms and sent to the other $m-1$ clients. $c_{i}$ holds the encrypted data, and no other client or servers can possess the complete $S_i$, achieving secure storage of gradients. During the decryption of $Encry_{P_i}(d_g)$, SMTFL does not require $c_i$ to perform decryption operations. The $s_a$ sends requests for the key shares of $S_i$ to all online clients. As long as at least $t$ clients response, the $s_a$ can recover $S_i$ and decrypt the poisoning gradients of malicious clients, even if $c_i$ or a few other clients are offline.  \hfill $\square$
\section{Experiment evaluation}
\label{Exp}

\subsection{Experimental setting}
\textbf{Environment}. We implement the experiments on a Linux server equipped with 1 NVIDIA RTX A6000 GPU. The experiments utilize the PyTorch framework, with CUDA version 11.4, Torch version 1.10.1+cu111, and the Python 3.6.13.
%%%%%%%%%%%%%%%%%%%%%%%%%%%%%%%%%%%%%%%%%
%是怎么设置出来 一大堆 客户端的,比如有200个clients,是怎么模拟出来的.我们肯定没200个真实设备跑
% 以及服务器,也就是中心聚合服务器
%%%%%%%%%%%%%%%%%%%%%%%%%%%%%%%%%%%%%%%%%

% Table generated by Excel2LaTeX from sheet 'Sheet5'
\begin{table*}[htbp]
  \centering
  \caption{Parameters Settings}
    \begin{tabular}{cccccccccc}
    \toprule
    Dataset & $num_{class}$ & Model & $num_{client}$ & $rate_{iid}$ & $thre_{eva}$ & $Epoch_{Local}$ & $Epoch_{Global}$ & Batchsize & Malicious Clients Proportions \\
    \midrule
    MNIST & 10    & CNN   & 150   & 0.5   & 6     & 20    & 30    & 100   & 25\% \\
    FMNIST & 10    & CNN   & 150   & 0.5   & 6     & 20    & 30    & 100   & 25\% \\
    CIFAR-10 & 10    & ResNet-18 & 150   & 0.5   & 6     & 50    & 30    & 100   & 25\% \\
    EMNIST & 47    & CNN   & 150   & 0.5   & 6     & 20    & 30    & 100   & 25\% \\
    \bottomrule
    \end{tabular}%
  \label{tab:addlabel}%
\end{table*}%

\textbf{Datasets and their partitioning}. We evaluate SMTFL in image classification tasks using four datasets: MNIST~\cite{726791}, Fashion-MNIST (FMNIST)~\cite{xiao2017fashion}, Extended MNIST (EMNIST)~\cite{7966217}, and CIFAR-10~\cite{Krizhevsky2009LearningML}. These datasets are commonly used as benchmark datasets in FL research, particularly under poisoning attack scenarios~\cite{nguyen2022flame, mo2021ppfl, kumari2023baybfedbayesianbackdoordefense, yin2021, konečný2017federatedlearningstrategiesimproving}. To assess SMTFL's performance across different data distributions, we introduce an independent and identically distributed (IID) rate, denoted as $rate_{iid}$, for partitioning the datasets, aligning with prior work~\cite{fang2020local, rieger2022deepsight, Fereidooni2023FreqFedAF}.

% We partition the clients into $n$ groups, and each group is assigned a group label $\{Glabel_1, Glabel_2, ..., Glabel_n\}$, for instance, the MNIST dataset, which contains 10 image classes, is partitioned into $n = 10$ groups. For each training data item with label $i$, it is assigned to a random client in the group $Glabel_i$ with a $rate_{iid}$, and to clients in other groups with a probability of $1 - rate_{iid}$. Within the same group, the data is uniformly distributed across clients. The $rate_{iid}$ controls the variation in data distribution across clients' training datasets. When $rate_{iid}=\frac{1}{n}$, the clients’ local training data is IID. Otherwise, the it is non-IID. Thus, $rate_{iid}$ is positively correlated with the degree of non-IID distribution of the training data.

\textbf{Models in FL}. To demonstrate the SMTFL's generalizability, we evaluate it by different combinations of datasets and models. The evaluation is conducted on the MNIST, EMNIST, and FMNIST datasets using a Convolutional Neural Network (CNN) and on the CIFAR-10 dataset using a Residual Neural Network (ResNet-18). 
 
\textbf{Baseline attacks}. We evaluate SMTFL against both poisoning attacks and gradient inversion attacks. For the poisoning attacks, we consider three common scenarios:
\begin{itemize}
\item Label Flipping Attack~\cite{shen2016auror}: Untrusted participants arbitrarily modify the labels of training data based on the gradients uploaded by the clients.

\item Random Update Attack~\cite{yin2021}: Malicious clients arbitrarily alter the uploaded gradients, leading to poor aggregation results for the global model and causing it to deviate from normal performance.

\item Projected Gradient Descent Attack (PGD)~\cite{madry2019deeplearningmodelsresistant}: Malicious clients utilize gradient information to generate adversarial samples that are visually similar to normal samples. However, these adversarial samples cause  global model to make incorrect predictions.
\end{itemize}

For the gradient inversion attacks, we consider  to implement the image reconstruction attacks~\cite{zhu2019deepleakagegradients}, where malicious participants attempt to reconstruct the training data based on the obtained gradients.

\textbf{Parameters}. Unless otherwise specified, we set the default number of clients to 150, with 25\% of them being malicious. 
%最大攻击次数为6次.
The data distribution is controlled by $rate_{iid}=0.5$. We explore the SMTFL's effectiveness against the aforementioned attacks under different proportions of malicious clients and data distributions. Table.~\ref{tab:addlabel} shows the default parameters settings, which will be employed unless otherwise mentioned. 
% The main notations are summarized in Table \ref{tab2}.
%%%%%%%%%%%%%%%%%%%%%%
% 生成小组梯度的部分,会有许多参数, 都会有一些影响,比如\varepsilon, 看一下.
%%%%%%%%%%%%%%%%%%%%%%

\textbf{Evaluated parameters}, including:
\begin{itemize}
    \item Performance changes of the global model. It includes the model's performance before and after poisoning attacks, and the performance changes after deploying SMTFL. The model's performance is measured by its accuracy.
    \item Defense against gradient inversion attacks. We compare the gradients and the reconstructed visual data before and after deploying SMTFL.
    \item The effectiveness of SMTFL in defending against poisoning attacks in the non-IID situation.
    \item Malicious client detection. We focus on identifying and removing malicious clients. We evaluate the SMTFL's ability to detect malicious clients and the number of epochs to detect malicious clients.
    \item False positive rate for malicious clients. We may mistakenly judge a honest client as a malicious client when evaluating a group of clients uniformly.
    
    \item Encryption overhead. We assess the time taken to generate shares in threshold encryption and the time required for gradient encryption and decryption.

    \item Storage overhead. As the storage server stores historical gradients from clients, we evaluate the storage space it needs to provide for each client.

\end{itemize}
\subsection{Experimental results}
%%%%%%%%%%%%%%%%%%%%%%%%%%%%%%%%%%%%%%%%%%
% 由于比较了2种模型 和 多种攻击手段,所以 下面的内容,都需要对每种进行介绍.  总数=模型数 * 攻击手段数,同类型的图,能整合到一张图最好.

% \textbf{全局模型的性能变化}. 全局模型在投毒攻击前后的性能变化. 设定客户端的数量.在不同的恶意客户端比例下,看全局模型的性能变化,作图. 为了让图看上去丰富点,可以一张图中有多种客户端数量 如{100,200,300,400,500},恶意比例{5\%,10\%,15\%,20\%,25\%,30\%, ..., 50\%}.具体这些值,你自己定,看怎么合适.

We perform a comprehensive evaluation of SMTFL based on the evaluated parameters outlined above. 

%针对上述评价指标,在此我们对SMTFL进行全面的评估.

\textbf{Effectiveness of SMTFL}. We conduct a thorough evaluation of SMTFL on four datasets using the two types of baseline attacks described above. 
% \textbf{SMTFL的有效性.}我们根据上文所述的两类基线攻击在四个数据集上对SMTFL进行全面评估.

\begin{figure}
    \centering
    \includegraphics[width=0.91\linewidth]{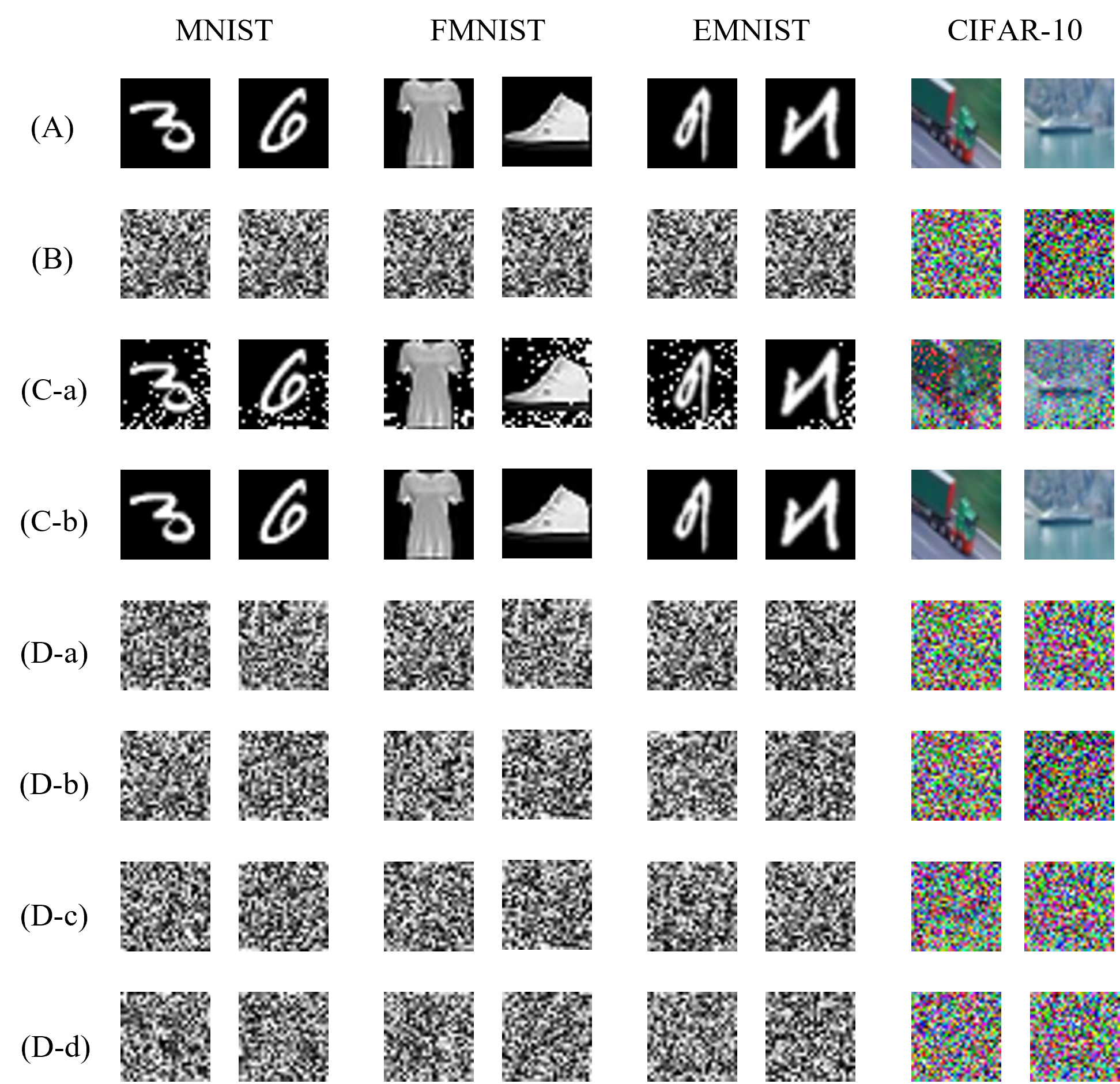}
    \caption{Effectiveness of SMTFL in defense against gradient inversion attacks on four datasets. (A) the original images; (B) the original attacked images; (C) the reconstructed images obtained through gradient inversion attacks during the 5-th and 10-th epochs of training, which means that very few epochs are needed to successfully reconstruct the images; (D) the reconstructed images obtained through gradient inversion attacks when SMTFL is deployed, where the images are reconstructed through $\{g_{A}^1$, $g_{A}^2+\varepsilon_A$, $g_{A^2,B}$, $g_{A, B, C}$\}.}
    \label{fig5}
\end{figure}

\begin{itemize}
    % \item \textbf{Defense against poisoning attacks}. We define untrusted clients that initiate poisoning attacks during the training process as "malicious participants," following the definition in related works~\cite{9546463, rieger2022deepsight}. 
    \item \textbf{Defense against poisoning attacks}. Assuming the fixed 150 clients, we consider scenarios where the number of malicious clients is at most half of the total clients (i.e., $\frac{m}{2}$). We evaluate SMTFL under various proportions of malicious clients to demonstrate its robustness. As shown in Table.\ref{tab3}, we report the change in global model accuracy before and after poisoning attacks for different malicious client proportions (that is, 5\%, 15\%, 25\%, 35\%, 45\%), along with the effectiveness of the SMTFL. Notably, we observe a slight increase in the model's accuracy in cases where the proportion of malicious clients is relatively low (around 5\%). We attribute this to a possible improvement in model generalization in the presence of minor attacks, which may enhance performance on downstream tasks to a certain extent.

    \item \textbf{Defense against gradient inversion attacks}. Semi-honest servers and clients can initiate gradient inversion attacks. Since SMTFL groups clients together, there are four stages in the gradient transmission process when clients upload gradient data to the server (e.g., $g_{A}^1$, $g_{A}^2+\varepsilon_A$, $g_{A^2,B}$, and $g_{A, B, C}$). Semi-honest servers and clients can attempt to access client's gradient data at each of these stages to perform gradient inversion attacks, aiming to reconstruct training data and compromise client's privacy. To assess the effectiveness of SMTFL against such attacks, we conducted experiments at each transmission stage. The results, shown in Fig.\ref{fig5}, demonstrate that servers and clients are unable to reconstruct client's training data from gradients acquired at any stage, thus protecting client's privacy.
\end{itemize}

% \textbf{1）防御投毒攻击.}我们定义投毒攻击过程中发起投毒攻击的不可信客户端为恶意参与方,在此我们与相关工作保持一致\cite{},在固定客户端数量为150的前提下,假设恶意参与方的数量至多为num_client/2,针对不同的恶意客户端比例对SMTFL进行评估,以展现其通用性.实验结果如表3所示,我们列举了在不同恶意客户端比例下,中毒攻击前后全局模型的精度变化,以及SMTFL方法的效果.特别的,我们观察到当系统中恶意参与方比例比较低（5%左右）时,模型受到了攻击之后测试精度存在小幅度上升的情况.我们认为在存在少量攻击的情况下,能在一定程度上提高模型的泛化性,更好的完成下游任务.

\begin{table*}[htbp]
  \centering
  \caption{The change in model accuracy before and after defense in SMTFL across different datasets, under various attack methods and malicious client proportions.}
    \begin{tabular}{ccccccccccc}
    \toprule
    \multirow{1.5}[4]{*}{Dataset} & \multirow{1.5}[4]{*}{Model} & \multirow{1.5}[4]{*}{No attack} & \multirow{1.5}[4]{*}{$num_{client}$} & \multirow{1.5}[4]{*}{Attack method} & \multirow{1.5}[4]{*}{Stage} & \multicolumn{5}{c}{Malicious client proportions \& Model accuracy} \\
\cmidrule{7-11}          &       &       &       &       &       & 5\% & 15\% & 25\% & 35\% & 45\% \\
    \midrule
    \multirow{6}[6]{*}{MNIST} & \multirow{6}[6]{*}{CNN} & \multirow{6}[6]{*}{95.23\%} & \multirow{6}[6]{*}{150} & \multirow{2}[2]{*}{PGD} & Attacked (\%) &94.87  &13.96  &9.80  &9.83  &9.77  \\
          &       &       &       &       & Defended (\%) & 95.64 & 96.28 & 96.65 & 96.22 & 95.56 \\
\cmidrule{5-11}          &       &       &       & \multirow{2}[2]{*}{Label Flip} & Attacked (\%) & 96.52  & 95.77  & 89.07  & 88.34  & 86.81  \\
          &       &       &       &       & Defended (\%) & 96.53 & 96.92 & 97.09 & 96.52 & 96.60 \\
\cmidrule{5-11}          &       &       &       & \multirow{2}[2]{*}{Random
Update} & Attacked (\%) & 95.84  & 95.29  & 88.42  & 59.71  & 25.60  \\
          &       &       &       &       & Defended (\%) & 95.85 & 96.11 & 95.92 & 95.93 & 95.59 \\
    \midrule
    \multirow{6}[6]{*}{CIFAR-10} & \multirow{6}[6]{*}{Resnet-18} & \multirow{6}[6]{*}{67.78\%} & \multirow{6}[6]{*}{150} & \multirow{2}[2]{*}{PGD} & Attacked (\%) & 77.59  & 73.68  & 71.17  & 57.30  & 53.34  \\
          &       &       &       &       & Defended (\%) & 78.13 & 76.78 & 76.95 & 64.06 & 61.90 \\
\cmidrule{5-11}          &       &       &       & \multirow{2}[2]{*}{Label Flip} & Attacked (\%) & 68.35  & 73.75  & 55.30  & 46.45  & 37.88  \\
          &       &       &       &       & Defended (\%) & 69.44 & 78.01 & 67.71 & 76.58 & 65.34 \\
\cmidrule{5-11}          &       &       &       & \multirow{2}[2]{*}{Random
Update} & Attacked (\%) & 73.26  & 49.81  & 29.38  & 20.60  & 16.30  \\
          &       &       &       &       & Defended (\%) & 73.31 & 73.11 & 73.85 & 66.57 & 66.80 \\
    \midrule
    \multirow{6}[6]{*}{FMNIST} & \multirow{6}[6]{*}{CNN} & \multirow{6}[6]{*}{88.19\%} & \multirow{6}[6]{*}{150} & \multirow{2}[2]{*}{PGD} & Attacked (\%) & 83.85  & 86.33  & 86.59  & 80.60  & 76.62  \\
          &       &       &       &       & Defended (\%) & 88.04 & 86.36 & 96.66 & 87.09 & 86.71 \\
\cmidrule{5-11}          &       &       &       & \multirow{2}[2]{*}{Label Flip} & Attacked (\%) & 86.65  & 81.23  & 46.51  & 28.44  & 11.15  \\
          &       &       &       &       & Defended (\%) & 87.77 & 88.04 & 87.49 & 87.51 & 87.54 \\
\cmidrule{5-11}          &       &       &       & \multirow{2}[2]{*}{Random
Update} & Attacked (\%) & 87.52  & 76.53  & 62.68  & 35.36  & 10.00  \\
          &       &       &       &       & Defended (\%) & 88.00 & 88.27 & 88.25 & 87.50 & 87.47 \\
    \midrule
    \multirow{6}[6]{*}{EMNIST} & \multirow{6}[6]{*}{CNN} & \multirow{6}[6]{*}{83.43\%} & \multirow{6}[6]{*}{150} & \multirow{2}[2]{*}{PGD} & Attacked (\%) & 75.78  & 64.27  & 35.23  & 27.97  & 18.36  \\
          &       &       &       &       & Defended (\%) & 80.75 & 82.70 & 82.22 & 81.91 & 80.92 \\
\cmidrule{5-11}          &       &       &       & \multirow{2}[2]{*}{Label Flip} & Attacked (\%) & 82.38  & 81.51  & 80.76  & 79.44  & 74.23  \\
          &       &       &       &       & Defended (\%) & 82.77 & 82.11 & 82.88 & 82.45 & 81.16 \\
\cmidrule{5-11}          &       &       &       & \multirow{2}[2]{*}{Random
Update} & Attacked (\%) & 82.82  & 80.12  & 35.22  & 27.97  & 18.36  \\
          &       &       &       &       & Defended (\%) & 83.03 & 82.11 & 82.22 & 81.91 & 80.92 \\
    \bottomrule
    \end{tabular}%
  \label{tab3}%
\end{table*}%

% \textbf{防御梯度翻转攻击。}我们定义梯度翻转攻击过程中发起恶意攻击的不可信服务器为恶意参与方。因SMTFL方法将客户端进行分组，因此，客户端将梯度数据上传到服务器的过程中共有四个阶段存在梯度数据的传输（如图3），恶意参与方会在各个阶段获取客户端梯度进行梯度翻转攻击，以窃取客户端训练数据从而侵犯隐私。基于此，我们对其各阶段进行实验，结果如图4所示，表明无论恶意参与方获得任一阶段的客户端梯度都无法重建其训练数据，进而保护了客户端隐私。

%(A) 表示原图，(B) 表示原始攻击图，(C) 分别表示第5轮和第10轮梯度重建攻击恢复的图像，(D) 表示我们的SMTFL系统，在客户端小组内产生的三个通信以及组聚合后与服务器的通信下，梯度泄露后的重建图像

 %我们在不同的参数配置下评估了SMTFL的性能，涵盖了多种数据分布和验证策略，旨在评估其鲁棒性和泛化能力。
\textbf{Parameter sensitivity analysis}. To assess SMTFL's robustness and generalization ability under different parameter settings, we evaluate the performance of SMTFL, including the model accuracy, the proportion of  correctly identified malicious clients, and the false positive rate for honest clients. 

\begin{figure*}
    \centering
    \includegraphics[width=1\linewidth]{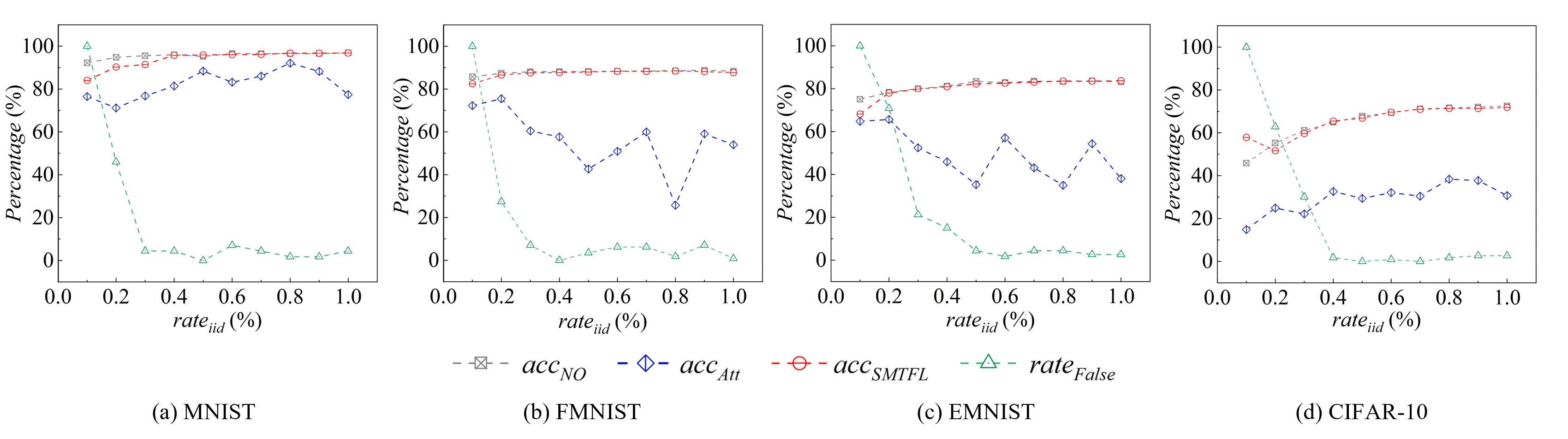}
    \caption{The model accuracy after deploying SMTFL ($acc_{SMTFL}$), model accuracy without attack ($acc_{No}$), model accuracy after being attack ($acc_{Att}$), and the false positive rate for honest clients ($rate_{False}$) across four datasets with different data distributions ($rate_{iid}$).}
    \label{fig6}
\end{figure*}

\begin{figure*}
    \centering
    \includegraphics[width=1\linewidth]{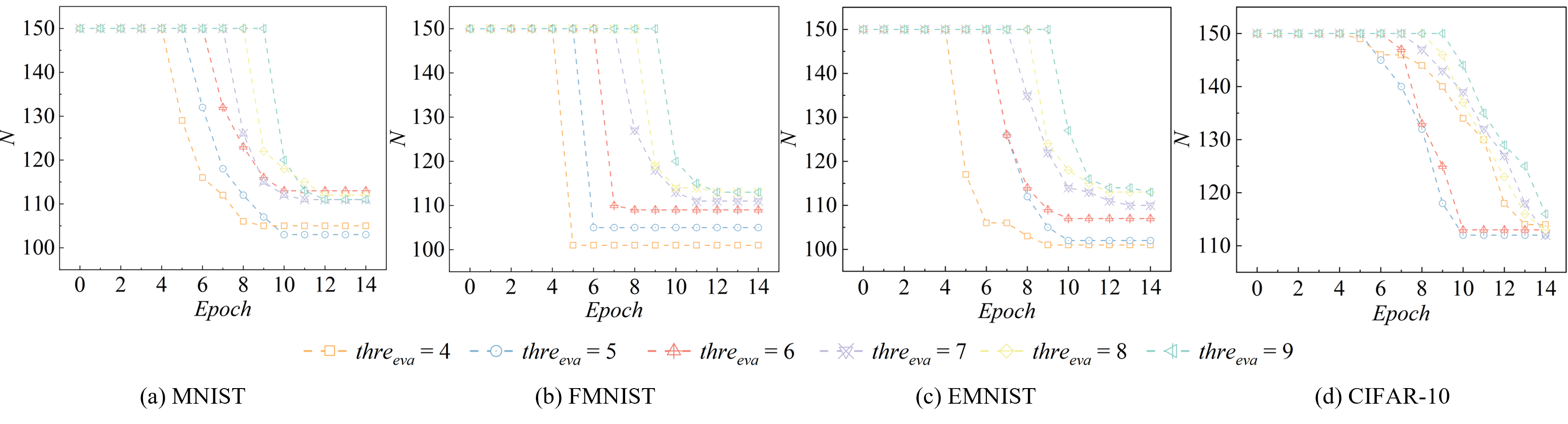}
    \caption{The trend of the number of clients participating in the FL system across different $thre_{eva}$ and epoch, on four datasets.}
    \label{fig7}
\end{figure*}

%数据分布的影响。我们在四个数据集上，针对不同比例的iid率评估SMTFL的性能，结果如图6所示。结果表明，SMTFL在不同的数据分布下都能准确定位恶意客户端，识别准确率（发现的/真正的）达97.3%以上。经过SMTFL方法，模型精度基本达到或略低于没有攻击时的模型精度。当iid率较低时（iid率<0.2），由于数据分布极为不均，导致FL系统中较多良性客户端被误伤，但SMTFL仍能有效抵御攻击。
\begin{itemize}
\item \textbf{Impact of $rate_{iid}$}. We evaluate the performance of SMTFL on four datasets with different $rate_{iid}$, as shown in Fig.\ref{fig6}. The results demonstrate that SMTFL accurately identifies malicious clients across different data distributions, achieving a proportion of correctly identified malicious clients (the ratio of detected to actual malicious clients) over 97.3\%. With the application of SMTFL, the model's accuracy remains largely on par with or slightly lower than that of the model without any attacks. When the $rate_{iid}$ is low (e.g., $rate_{iid} \leq 0.2$), the highly imbalanced data distribution leads to a higher number of honest clients being misclassified as malicious. Nevertheless, SMTFL effectively mitigates the attacks and maintains strong robustness.

\item \textbf{Impact of $thre_{eva}$}. To enhance the robustness of FL systems, SMTFL effectively removes malicious clients $c_{mali}$ by setting a threshold $thre_{eva}$. This paper conducts comprehensive experiments on four datasets to evaluate the model's performance and computational overhead under different threshold settings $thre_{eva}$. The results on model performance are presented in Table.\ref{thre}. The findings demonstrate that regardless of the specific threshold setting, the SMTFL can efficiently identify and remove malicious clients, thereby restoring model accuracy, with the detection accuracy ($acc_{loc}$) for malicious clients  is upper  95\%. As the  $thre_{eva}$ increases, the false positive rate ($rate_{False}$) for honest clients generally decreases, a trend particularly notable in the FMNIST dataset. For the other three datasets, although some fluctuations in false positives were observed, the overall trend aligns with that of the FMNIST dataset. However, it is noteworthy that increasing the threshold is not always the optimal choice, as higher threshold values can prolong the persistence of malicious clients in the system, significantly increasing computational overhead. We further analyze, under different threshold settings across four datasets, the number of clients in the system after continuously detecting and removing malicious clients in each epoch. The results, shown in Fig.~\ref{fig7}, indicate that higher threshold values slow down the removal of malicious clients, thereby adding to the system’s computational burden. Setting the threshold requires balancing performance and computational overhead to achieve the optimal trade-off between model accuracy and resource consumption. When $thre_{eva}=6$, the model achieves the optimal recovery accuracy with minimal adverse effects on honest clients.
\end{itemize}
%惩罚因子的影响。为提升联邦学习（Federated Learning, FL）系统的鲁棒性，SMTFL通过设定阈值 x 来有效剔除恶意客户端c。本文在四个数据集上针对不同阈值x 进行了深入的实验，评估了模型在性能与计算开销方面的表现。模型性能的实验结果如表x所示。结果表明，无论阈值的具体设定如何，所提出的模型均能高效识别并剔除攻击者，从而恢复模型的精度，且对恶意客户端的检测精度可达95%。随着阈值 x 的增加，模型对良性客户端的误检率整体呈下降趋势，这在FMNIST数据集上的表现尤为显著。而对于其他三个数据集，尽管存在一定的误检波动，但总体趋势与FMNIST数据集相符。然而，值得注意的是，阈值的增大并不总是最佳选择，因为较高的阈值会导致恶意客户端在系统中持续存在的时间延长，从而显著增加系统的计算开销。
%因此，本文分析了四个数据集中不同阈值设置下每一轮参与训练的客户端数量。实验结果如图x所示，较高阈值会减缓攻击者从系统中移除的速度，这在一定程度上增加了系统的计算负担。在设置阈值时需在性能和计算开销之间进行权衡，以实现性能与资源消耗的最优平衡。当阈值为6时，模型的恢复精度达到最优，对模型的误伤相对较小。

\begin{table*}[htbp]
  \centering
  
  \caption{SMTFL's performance on four datasets under different threshold settings, including model accuracy ($acc_{SMTFL}$), malicious client localization accuracy ($acc_{loc}$), and false positive rate ($rate_{False}$) for honest clients.}
  \resizebox{\linewidth}{!}{
    \begin{tabular}{ccccccccccccc}
    \toprule
    \multirow{2}[2]{*}{$thre_{eva}$} & \multicolumn{3}{c}{MNIST} & \multicolumn{3}{c}{FMNIST} & \multicolumn{3}{c}{EMNIST} & \multicolumn{3}{c}{CIFAR-10} \\
    \cmidrule(r){2-4}\cmidrule(lr){5-7}\cmidrule(lr){8-10}\cmidrule(lr){11-13}

    & $acc_{SMTFL}$   & $acc_{loc}$ & $rate_{False}$   & $acc_{SMTFL}$   & $acc_{loc}$ & $rate_{False}$  & $acc_{SMTFL}$& $acc_{loc}$ & $rate_{False}$   & $acc_{SMTFL}$   & $acc_{loc}$ & $rate_{False}$ \\
    \midrule
    4     & 96.16\%  & 100.00\%  & 7.08\%  & 86.82\%  & 100.00\%  & 10.62\%  & 82.12\%  & 100.00\%  & 10.62\%  & 65.46\%  & 95.00\%  & 0.88\%  \\
    5     & 96.35\%  & 100.00\%  & 8.85\%  & 87.06\%  & 100.00\%  & 7.08\%  & 81.94\%  & 100.00\%  & 9.73\%  & 66.93\%  & 100.00\%  & 0.88\%  \\
    6     & 95.92\%  & 100.00\%  & 0.00\%  & 87.95\%  & 100.00\%  & 3.54\%  & 83.43\%  & 100.00\%  & 4.42\%  & 67.78\%  & 100.00\%  & 0.00\%  \\
    7     & 96.00\%  & 100.00\%  & 1.77\%  & 87.90\%  & 100.00\%  & 1.77\%  & 82.40\%  & 100.00\%  & 2.65\%  & 67.11\%  & 100.00\%  & 0.88\%  \\
    8     & 96.03\%  & 100.00\%  & 0.88\%  & 88.20\%  & 100.00\%  & 0.00\%  & 82.71\%  & 100.00\%  & 0.00\%  & 66.97\%  & 100.00\%  & 1.00\%  \\
    9     & 96.10\%  & 100.00\%  & 1.77\%  & 88.16\%  & 100.00\%  & 0.00\%  & 81.91\%  & 100.00\%  & 0.88\%  & 66.71\%  & 97.30\%  & 1.77\%  \\

    \bottomrule
    \end{tabular}%
    }
  \label{thre}%
\end{table*}%

%分组策略有效性
%SMTFL中的分组策略不仅能够有效保护客户端隐私，防止梯度泄露，还对模型训练有一定优势。为了更好地验证我们分组策略在提升模型性能和加快收敛速度方面的优势，我们在MNIST数据集上进行了对比分析。具体而言，我们比较了单个客户端训练并上传梯度至服务器，与分组后聚合梯度上传至服务器这两种方案在训练轮数变化中的模型精度趋势。结果如图x所示，实验结果表明，分组后的梯度聚合比单个客户端的梯度上传具有更好的收敛性能。此外，分组后的客户端聚合比单个客户端聚合更能有效地提升模型性能。通过使用不同比例的数据集来训练相同的模型并比较其性能提升效果，我们发现，当一个包含更多数据的客户端参与联邦学习（FL）系统时，模型的泛化能力得到了显著提升。同样地，我们的分组策略通过梯度聚合证明了它在提升模型泛化性能方面的有效性。而单个数据量较少的客户端在训练时可能会导致梯度偏离正常的更新范围，从而影响模型的整体表现。

\begin{figure}
    \centering
    \includegraphics[width=0.7\linewidth]{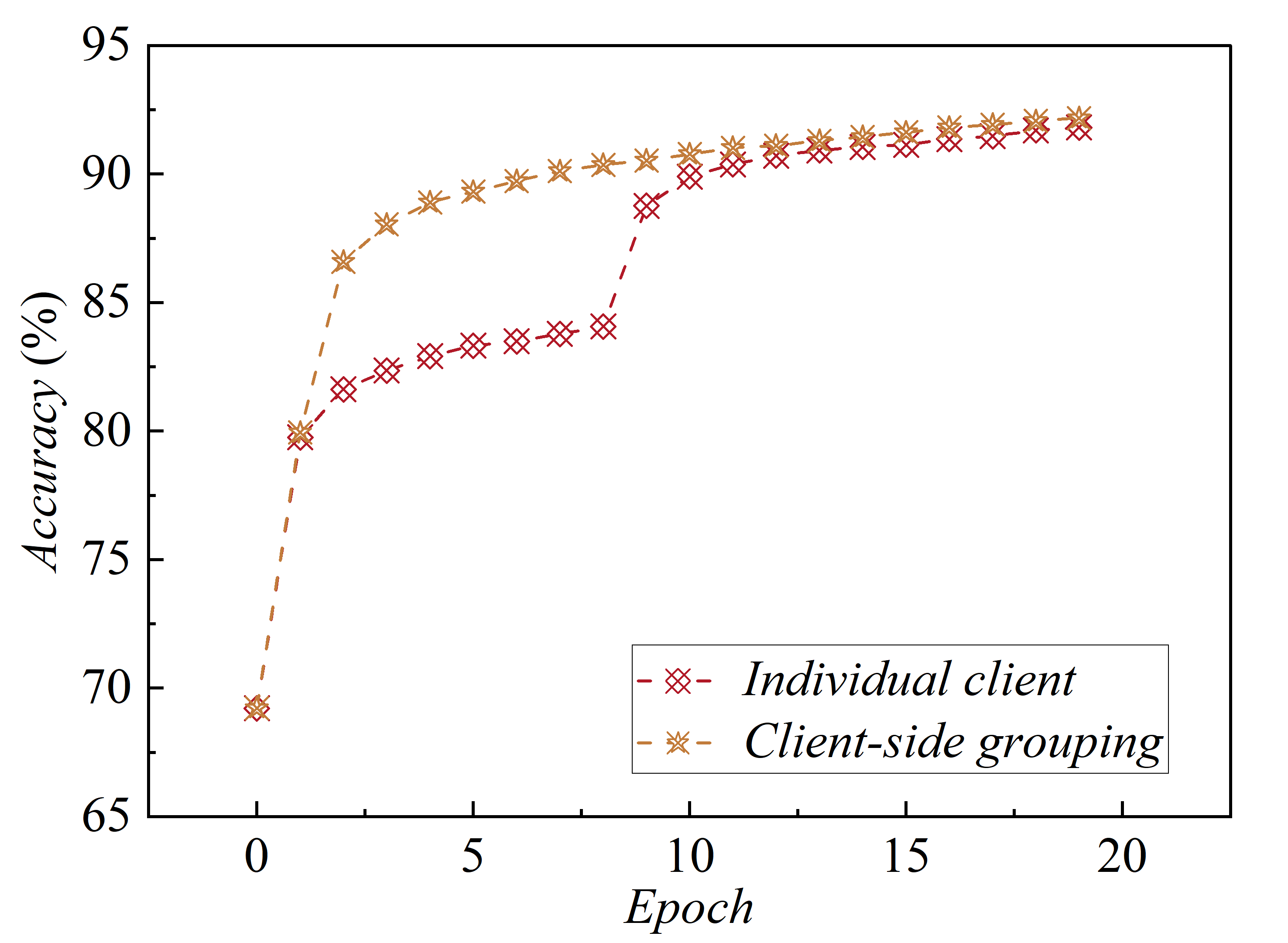}
    \caption{The trend of model accuracy with respect to the number of rounds, under the methods of individual client gradient aggregation and grouped gradient aggregation.}
    \label{fig8}
\end{figure}

\textbf{Grouping strategy provides extra advantages for model training}. SMTFL's grouping strategy  not only effectively protects client privacy and prevents gradient leakage but also offers certain advantages for model training. To better demonstrate the benefits of our grouping strategy in improving model performance and convergence speed, we conducted a comparative analysis on the MNIST dataset. Specifically, we compared the trend of model accuracy over training rounds between two approaches: training with individual client gradient uploads to the server and training with grouped client gradients aggregated and uploaded to the server. The results, shown in Fig.\ref{fig8}, indicate that gradient aggregation from grouped clients leads to superior convergence performance compared to individual client uploads. Furthermore, grouped client aggregation enhances model performance more effectively than individual client aggregation. By training the same model on datasets of varying sizes and comparing their performance improvements, we observed that when a client with a larger dataset participates in the FL system, the model's generalization ability is enhanced. Similarly, our grouping strategy, through gradient aggregation, has proven to be an effective method for improving the model's generalization performance. In contrast, clients with smaller datasets may experience gradient updates that deviate from the normal range, potentially affecting the overall model's performance.

% 计算存储与加密开销
% 在SMTFL系统中，我们引入了门限加密的理念，针对模型训练过程中生成的梯度数据进行加密存储。通过实验验证，我们分析了该加密方法的时间开销，实验结果如表x所示。我们的加密方案在确保数据安全性的同时，不增加计算负担，并能够在计算资源有限的环境中保持较高效率。在不同规模的数据集和多种模型结构下，我们的加密方法表现出了较强的可扩展性。

\begin{table}[htbp]
  \centering
  \caption{The storage overhead of the model and the encryption/decryption time overhead across four datasets.}
  % \resizebox{\linewidth}{!}{
    \begin{tabular}{ccccccc}
    \toprule
    Dataset & $Space_{sto}$  & $m$ &  $t$ & $T_{enc}$ (s) & $T_{dec}$ (s) & $T_{gks}$ (s) \\
    \midrule
    MNIST & 0.51 MB  & 149   & 90    & 0.0022 & 0.0056 & 0.0044 \\
    FMNIST & 0.69 MB  & 149   & 90    & 0.0024 & 0.0057 & 0.0042 \\
    CIFAR-10 & 256.37 MB  & 149   & 90    & 0.8307 & 0.8215 & 0.0042 \\
    EMNIST & 2.01 MB & 149   & 90    & 0.0078 & 0.0084 & 0.0042 \\
    \bottomrule
    \end{tabular}%
    % }
  \label{time}%
\end{table}%

\textbf{Encryption and storage overhead}. In the SMTFL system, we implement a threshold encryption mechanism to encrypt and store the gradient data generated during the model training process. The number of clients $m$ is 150, and the threshold $t$ is 90.  We assess the time overhead associated with this encryption method, which includes the time for encryption, decryption, and generation of key shares, that is, $T_{enc}$, $T_{dec}$, and $T_{gks}$. Concurrently, we evaluate the storage space required by the storage server for each client. The results are detailed in Table.~\ref{time}. Our encryption method ensures data security without significantly increasing the computational burden and maintains high efficiency, even in environments with constrained resources. The method exhibits the scalability across datasets of varying sizes and different model architectures.

% \textbf{Encryption and storage overhead} In the SMTFL system, we introduce the concept of threshold encryption to encrypt and store gradient data generated during the model training process. We validated the time overhead of this encryption method through experiments, and the results are presented in Table.\ref{time}. Our encryption approach ensures data security without increasing computational burden, while maintaining high efficiency even in resource-constrained environments. The method demonstrates strong scalability across datasets of varying sizes and different model architectures.
% 值得声明的是,在实验中，我们模拟了150个客户端组成联邦学习系统，但未考虑实际的通信开销。与传统的联邦学习方法相同，我们的方法未引入额外的通信开销。
% It is worth noting that we simulated a federated learning system consisting of 150 clients in the experiments, without considering the actual communication overhead. Similar to conventional federated learning methods, our approach does not introduce any additional communication overhead.

%%%%%%%%%%%%%%%%%%%%%%%%%%%%%%%%%%%%%%%%%%%%%%%

%%%%%%%%%%%%%%%%%%%%%%%%%%%%%%

\subsection{Comparison with related work}
We compare SMTFL with related works on the following aspects: defense against poisoning attacks, gradient privacy protection, the ability to locate malicious clients, model performance robustness, the requirement for clean data on the server, and limitations, as presented in Table.~\ref{relate}.

Some existing studies~\cite{cao2020fltrust, cao2023fedrecover} assume that servers are trustworthy and pose no threat to the privacy of stored data. Some works even require servers to access a small amount of public training data from each client, which can be challenging to meet and increases the server's workload. Therefore, this assumption is often inadequate in practical scenarios. In this paper, we consider that both the server and clients are untrustworthy, that is, the FL system has no trusted participants. In SMTFL, the server cannot obtain precise gradients for all clients, and the decryption key for stored information is not controlled by a single party. Instead, we employ threshold encryption, which enhances the system's security and stability by ensuring that no single entity has access to the full decryption key.

Most existing~\cite{cao2020fltrust, 10296389, DBLP:journals/corr/abs-2009-03561, Fereidooni2023FreqFedAF} studies concentrate on either defending against poisoning attacks or protecting gradient privacy, with few addressing both issues concurrently. SMTFL defend against poisoning attacks and gradient inversion attacks, thereby protecting the model correctness and the pravicy of training data. Furthermore, among the works that focus on poisoning attacks, the majority aim to detect malicious gradients. The construction of a trustworthy FL system, which involves removing malicious clients from the system, is not a primary consideration for them. In contrast, SMTFL builds a secure FL system by evaluating clients and eliminating malicious ones from the system.

We also compare the storage requirements per client. Specifically, prior to detecting malicious clients, the server retains historical gradients from the training process of the global model, such as each client's gradient updates for every epoch. Utilizing this stored historical gradients, the server estimates the gradient of malicious clients for each epoch when removing the poisoning gradient from the global model via the FUL. In contrast to the method described in~\cite{cao2023fedrecover}, which demands an average of 2 GB of additional storage per client for training on MNIST and FMNIST datasets, SMTFL requires less than 0.69 GB per client. Meanwhile, we conduct experiments on more datasets, including the complex CIFAR-10 dataset, which requires 0.25 GB of storage space per client.

% While achieving exceptional performance and robustness, the SMTFL model also excels in terms of storage overhead. Specifically, before detecting malicious clients, the server stores historical information during the training of a compromised global model, including the global model and client model updates for each round. During the recovery process, the server uses the stored historical information to estimate the client model updates for each round. Compared to method~\cite{cao2023fedrecover}, which requires an average of 2GB of additional storage per client, the SMTFL model only requires an average of 0.25GB per client.

% 相关工作就是FLtrust那些吧,感觉一些指标能作图,一些就只能定性描述比较了.
%%%%%%%%%%%%%%%%%%%%%%%%%%%%%%%%%%%%%%%%%%%%%%%%
% 展开每个指标进行文字介绍.
% 比较的话,其实也是比较上面的评价指标,如果想比较的更细致点的话.
% 比较结果的展示,应该不只是有表,也会有一些图,如不同方法在同一参数设置下,它们的结果差异展示.
%%%%%%%%%%%%%%%%%%%%%%%%%%%%%%%%%%%%%%%%%%%%%%%%
\begin{table*}[htbp]
\caption{Comparison among SMTFL and related work}\label{relate}
\begin{center}
\begin{tabular}{m{60pt}m{35pt}m{30pt}m{60pt}m{40pt}m{40pt}m{160pt}}
\hline
\makecell[c]{Work}
&{\makecell[c]{Poisoning} 

\makecell[c]{attack}} 
&{\makecell[c]{Gradient}

\makecell[c]{privacy}}
&{\makecell[c]{Malicious clients} 

\makecell[c]{location}} 
& {\makecell[c]{System} 

\makecell[c]{robustness}} 
& {\makecell[c]{Clean data}

\makecell[c]{in server}} 
& \makecell[c]{Limitation}\\
\hline

\makecell[c]{FLTrust~\cite{cao2020fltrust}}
&\makecell[c]{\checkmark} 
&{\makecell[c]{$\times$}} 
& \makecell[c]{\checkmark}
&{\makecell[c]{$\times$}}
&{\makecell[c]{\checkmark}} 
& {Server is trusted and need to collect and train on a small amount of publicly available data.}\\

\makecell[c]{FLAIRS \cite{10296389}}
&\makecell[c]{$\times$} 
&\makecell[c]{\checkmark} 
& \makecell[c]{\checkmark}
&{\makecell[c]{\checkmark}}
&{\makecell[c]{$\times$}} 
& {It depends on the trusted execution environment, and it does not support the non-iid data.}\\

\makecell[c]{Naseri et al. \cite{DBLP:journals/corr/abs-2009-03561}}
&\makecell[c]{$\times$} 
&\makecell[c]{\checkmark} 
&\makecell[c]{$\times$} 
&\makecell[c]{$\times$} 
&\makecell[c]{$\times$} 
& {It only focuses on the gradient's privacy and degrades the model's performance.}\\

\makecell[c]{EIFFeL~\cite{roy2022eiffel}}
&\makecell[c]{\checkmark} 
&\makecell[c]{\checkmark} 
&\makecell[c]{$\times$}
&{\makecell[c]{$\times$}}
&{\makecell[c]{$\times$}} 
&{Communication and computation costs are high, may not be available in real applications. It is limited by the number of malicious clients.}\\

\makecell[c]{FreqFed~\cite{Fereidooni2023FreqFedAF}}
&{\makecell[c]{\checkmark}}
&\makecell[c]{$\times$} 
& {\makecell[c]{\checkmark}}
&{\makecell[c]{$\times$}}
&{\makecell[c]{$\times$}} 
&{
It only focus on the poisoning attacks and is applicable to image data.}\\
\hline
\makecell[c]{SMTFL}
&\makecell[c]{\checkmark} 
&\makecell[c]{\checkmark} 
&\makecell[c]{\checkmark}
&\makecell[c]{\checkmark}
&\makecell[c]{$\times$}
&\makecell[c]{—}\\
\hline

\end{tabular}
\end{center}
\end{table*}

\section{Related work}
\label{re}
% 每部分需要标注多个参考文献,即使没有细节它们.参考文献尽可能的新
For poisoning and gradient inversion attacks, current defense strategies mainly include the following three ideas.  

\textbf{Anomaly detection-based methods}. They defend against poisoning attacks by identifying and excluding the gradient updates of malicious clients. However, they are often limited by filtering strategies and data distribution assumptions. In term of filtering strategies, the similarity of gradients from different clients can be leveraged to filter a small number of malicious clients, but it is susceptible to collusion attacks. Fung et al.~\cite{fung2020limitations} employ the KMeans method to achieve filtering, but this method has high computational complexity, and malicious clients can circumvent defenses by submitting multiple backdoored samples. Auror method~\cite{shen2016auror} runs solely on the server and has a lower performance overhead, but malicious clients can exploit the updates diversity to enhance the collusion attacks. In term of data distribution assumptions, existing methods often rely on specific assumptions. Andreina et al.~\cite{9546463} and Cao et al.~\cite{cao2020fltrust} assume that the server has access to clients' training data, which violates client's data privacy. Variations in data distribution (such as IID or non-IID) can cause differences in the clients' gradients. FLAIRS~\cite{10296389} and FreqFed~\cite{Fereidooni2023FreqFedAF} identify the abnormal gradients of malicious clients through anomaly detection algorithms (e.g., cosine distance and HDBSCAN), where these gradients are assumed to contain obvious outliers. 

\textbf{Differential privacy-based methods}. Noise is added to the clients' gradients to prevent attackers from obtaining precise gradients, thereby avoiding the data reconstruction and mitigating negative impact on the performance of global model. Compared to anomaly detection methods, DP considers the risk of data leakage during gradient transmission. However, since the server cannot access clients' training data, it cannot accurately analyze and eliminate the effects of noise in gradient aggregation, which degrades the performance of global model. Naseri et al.~\cite{DBLP:journals/corr/abs-2009-03561} implement a collaboration between local and central DP, where clients add the noise to gradients while the server uses DP aggregation algorithms for gradient aggregation. As the number of malicious clients increases, it is more challenging for the server to analyze the noise. Based on the parameter clipping and Gaussian noise adding, McMahan et al.~\cite{DBLP:journals/corr/abs-1710-06963} modify the gradients to limit gradient sizes and protect data privacy, which has expensive computational costs. FLAME~\cite{nguyen2022flame} combines the filtering of outlier detection, model clipping, and noise addition, but its privacy guarantees are only applicable to semi-honest server that adheres to the Secure Multi-Party Computation (SMPC) protocol.

\textbf{Secure Aggregation-Based Methods}. The goal is to achieve that the performance of global model is not significantly affected, even if some clients upload incorrect gradients. The server will not detect malicious clients during gradient aggregation. Instead, it employs specific aggregation strategies to enhance the robustness of global model. Multi-Krum method~\cite{10.5555/3294771.3294783} repeatedly removes these gradient updates that are far from the geometric median of all updates. Trimmed Mean method~\cite{yin2021} removes the maximum / minimum updates and computes the average of the remaining values. Although these methods mitigate the impact of incorrect gradients to some extent, the performance of global model still significantly declines as the number of attackers increases. Pasquini et al.~\cite{10.1145/3548606.3560557} claim that current secure aggregation-based FL achieves a ``false sense of security", i.e., it merely addresses superficial privacy protection without defending against client-side attacks. To achieve privacy protection and model correctness, Chowdhury et al.~\cite{roy2022eiffel} split and aggregate the clients' gradients through the Shamir threshold secret sharing scheme and non-interactive proofs. This method has high computational complexity, making it challenging to balance the model robustness with the computational efficiency.

Overall, existing methods mitigate gradient inversion attacks and poisoning attacks, but they exhibit the limitations in complex attack scenarios and rarely address both security issues simultaneously. Against this background, the SMTFL offers a novel solution. Our proposed method not only effectively counters both gradient inversion attacks and poisoning attacks but also demonstrates outstanding robustness and stability in complex attack scenarios. Our method enables secure model training in FL involving untrusted participants, providing new perspectives and insights for addressing similar challenges.

\section{Discussion and Conclusion}
\label{Dis}
 % \textbf{一些问题的思考过程,来表征我们方案是经过深思熟虑的,合理的}
In FL system, this paper introduces an approach called SMTFL to achieve our security goals: preventing participants from inferring clients' training data from their gradients, detecting malicious clients that execute poisoning attacks, and ensuring the correctness of the global model. Regarding privacy protection, we have eschewed the use of noise that could compromise model performance to obfuscate gradients. Instead, we have implemented a system of checks and balances among groups of clients to prevent collusion, ensuring that no participant can access the gradients of others. In terms of detecting and defending against poisoning attacks, we have not relied on a trusted participant, nor have we required the server to collect public training data from each client. We determine the aggregation of a gradient into the global model based on the performance changes of the global model, thereby circumventing the collusion attacks encountered by strategies that depend on the gradient direction uploaded by the majority of clients. We have evaluated and demonstrated the effectiveness of SMTFL through image classification tasks. We are surprised to find that SMTFL slightly improves the performance of the global model. Based on our current research, \textbf{we will attempt the following work in the future}:
\begin{itemize}
    \item We will measure to reduce the impact on mistakenly targeted clients, such as optimizing group partitioning strategies and unified evaluation strategies. 
    \item Without disclosing the gradients, we will mitigate the impact of poisoning gradients on the global model by exploring two ideas: 1) Preventing the poisoning gradients from being aggregated into the global model. 2) Although poisoning gradients have been aggregated into the global model, simplifying the gradients' storage, encryption, and decryption processes on the storage server, and efficiently decrypting and utilizing FUL to reduce the impact of poisoning gradients on the global model. 
    \item Model training methods that support multimodal data are of great importance. We believe that the grouping strategy of SMTFL is applicable to training with different types of data. We will adjust SMTFL to accommodate more complex real-world scenarios.
\end{itemize}

Currently, research in fields such as large language models~\cite{zhao2023survey}, Embodied AI~\cite{yang2024holodeck}, and edge intelligence~\cite{zhao2024feashare} continues to gain momentum, leading to a substantial increase in data and model training demands. Third-party clients and aggregation servers are emerging rapidly to participate in FL applications to meet these demands. Correspondingly, ensuring users receive correct global models and protecting the client's data privacy have become more critical and urgent.

\section*{Acknowledgment}

We would like to thank the anonymous reviewers for their insightful comments and suggestions.

% \ifCLASSOPTIONcaptionsoff
%   \newpage
% \fi

\bibliographystyle{IEEEtran}
\bibliography{ref} %bibfile_name

\end{document}